\documentstyle[12pt]{article}

\input epsf

\def\WIMPZILLA{{\sc wimpzilla}}
\def\WIMPZILLAS{{\sc wimpzillas}}
\def\WIMP{{\sc wimp}}
\def\WIMPS{{\sc wimps}}
\def\LTE{{\sc lte}}

\def\ktilde{{{\tilde{k}}}}
\def\etatilde{{{\tilde{\eta}}}}
\def\simlt{\stackrel{<}{{}_\sim}}
\def\simgt{\stackrel{>}{{}_\sim}}
\newcounter {subsubsubsection}[subsubsubsection]
\renewcommand{\thesubsubsubsection} {\thesubsubsection.\alph{subsubsubsection}}
\def\subsubsubsection#1{\vspace*{12pt} 
\centerline{\thesubsubsubsection{\em  #1} \addtocounter{subsubsubsection}{1}} }

\begin{document}
\begin{center}
{\Large \bf Dynamics of the Inflationary Era}\\
\vspace*{0.7cm}
{\bf Edward W. Kolb} \\
\vspace*{ 0.5cm}
{\it NASA/Fermilab Astrophysics Center \\ 
Fermi National Accelerator Laboratory, Batavia, Illinois~~60510-0500\\
and\\
Department of Astronomy and Astrophysics, Enrico Fermi Institute\\
The University of Chicago, Chicago, Illinois~~60637-1433}
\end{center}
\vspace*{0.1cm}
\begin{quote}
There is very strong circumstantial evidence that there was an
inflationary epoch very early in the history of the universe.  In this
lecture I will describe how we might be able to piece together some
understanding of the dynamics during and immediately after the
inflationary epoch.
\end{quote}

\section{Introduction}

We live in a very large, very old, and nearly (perhaps exactly)
spatially flat universe.  The universe we observe is, at least on
large scales, remarkably homogeneous and isotropic.  These attributes
of our universe can be ascribed to a period of very rapid expansion,
or inflation, at some time during the very early history of the
universe \cite{alan}.

A sufficiently long epoch of primordial inflation leads to a
homogeneous/isotropic universe that is old and flat.  The really good
news is that very many reasonable models have been proposed for
inflation.\footnote{Perhaps a more accurate statement is that there
are many models that seemed reasonable to the people who proposed them
at the time they were proposed.}  In some ways, inflation is generic.
That is also the really bad news, since we would like to use the early
universe to learn something about physics at energy scales we can't
produce in terrestrial laboratories.  We want to know more than just
that inflation occurred, we want to learn something of the dynamics of
the expansion of the universe during inflation.  That may tell us
something about physics at very high energies.  It would also allow us
to restrict the number of inflation models.  If we can differentiate
between various inflation models, then inflation can be used as a
phenomenological guide for understanding physics at very high
energies.

Oldness, flatness, and homogeneity/isotropy only tell us the minimum
length of the inflationary era.  They are not very useful instruments
to probe the dynamics of inflation.  If that is our goal, we must find
another tool.  In this lecture I will discuss two things associated
with inflation that allow us to probe the dynamics of inflation:
perturbations and preheating/reheating.

While the universe is homogeneous and isotropic on large scales, it is
inhomogeneous on small scales.  The inhomogeneity in the distribution
of galaxies, clusters, and other luminous objects is believed to
result from small seed primordial perturbations in the density field
produced during inflation.  Density perturbations produced in
inflation also lead to the observed anisotropy in the temperature of
the cosmic microwave background radiation.  A background of primordial
gravitational waves is also produced during inflation.  While the
background gravitational waves do not provide the seeds or influence
the development of structure, gravitational waves do lead to
temperature variations in the cosmic microwave background radiation.
If we can extract the primordial density perturbations from
observations of large-scale structure and cosmic microwave background
radiation temperature fluctuations, then we can learn something about
the dynamics of inflation.  If we can discover evidence of a
gravitational wave background, then we will know even more about the
dynamics of the expansion rate during inflation.

Inflation was wonderful, but all good things must end.  The early
universe somehow made the transition from an inflationary phase to a
radiation-dominated phase.  Perhaps there was a brief matter-dominated
phase between the inflation and radiation eras.  In the past few years
we have come to appreciate that some interesting phenomena like phase
transitions, baryogenesis, and dark matter production can occur at the
end of inflation.  Perhaps by studying how inflation ended, we can
learn something of the dynamics of the universe during inflation.

\section{Perturbations Produced During Inflation}

One of the striking features of the cosmic background radiation (CBR)
temperature fluctuations is the growing evidence that the fluctuations
are acausal.\footnote{Exactly what is meant by acausal will be explained
shortly.  Acausal is in fact somewhat of a misnomer since, as we shall
see, inflation produces ``acausal'' perturbations by completely causal
physics.} The CBR fluctuations were largely imprinted at the time of
last-scattering, about 300,000 years after the bang.  However, there
seems to be fluctuations on length scales much larger than 300,000 light
years!\footnote{Although definitive data is not yet in hand, the
issue of the existence of acausal perturbations will be settled very
soon.}  How could a causal process imprint correlations on scales
larger than the light-travel distance since the time of the bang?  The
answer is inflation.

In order to see how inflation solves this problem, first consider the
evolution of the Hubble radius with the scale factor
$a(t)$:\footnote{Here and throughout the paper ``RD'' is short for
radiation dominated, and ``MD'' implies matter dominated.}
\begin{equation}
R_H\equiv H^{-1} = \left(\frac{\dot{a}}{a}\right)^{-1} 
\propto \rho^{-1/2}\propto \left\{\begin{array}{ll} a^2 & \mbox {(RD)}\\ 
        a^{3/2} & \mbox {(MD)}. \end{array} \right. 
\end{equation}
In a $k=0$ matter-dominated universe the age is related to $H$ by $t =
(2/3)H^{-1}$, so $R_H=(3/2)t$.  In the early radiation-dominated
universe $t=(1/2)H^{-1}$, so $R_H=2t$.

\begin{figure}[t]
\centering
\leavevmode\epsfxsize=220pt  \epsfbox{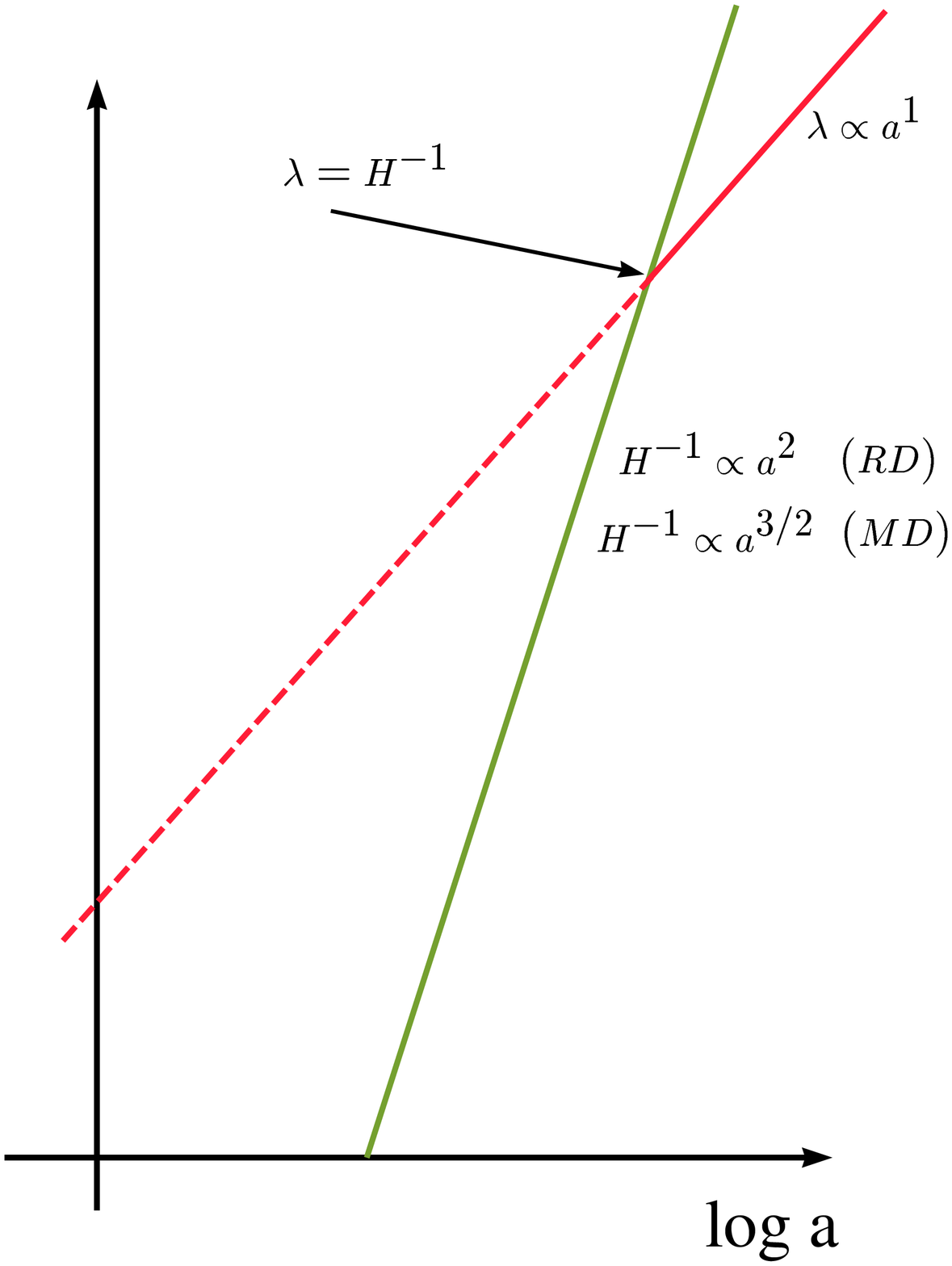}
\leavevmode\epsfxsize=220pt  \epsfbox{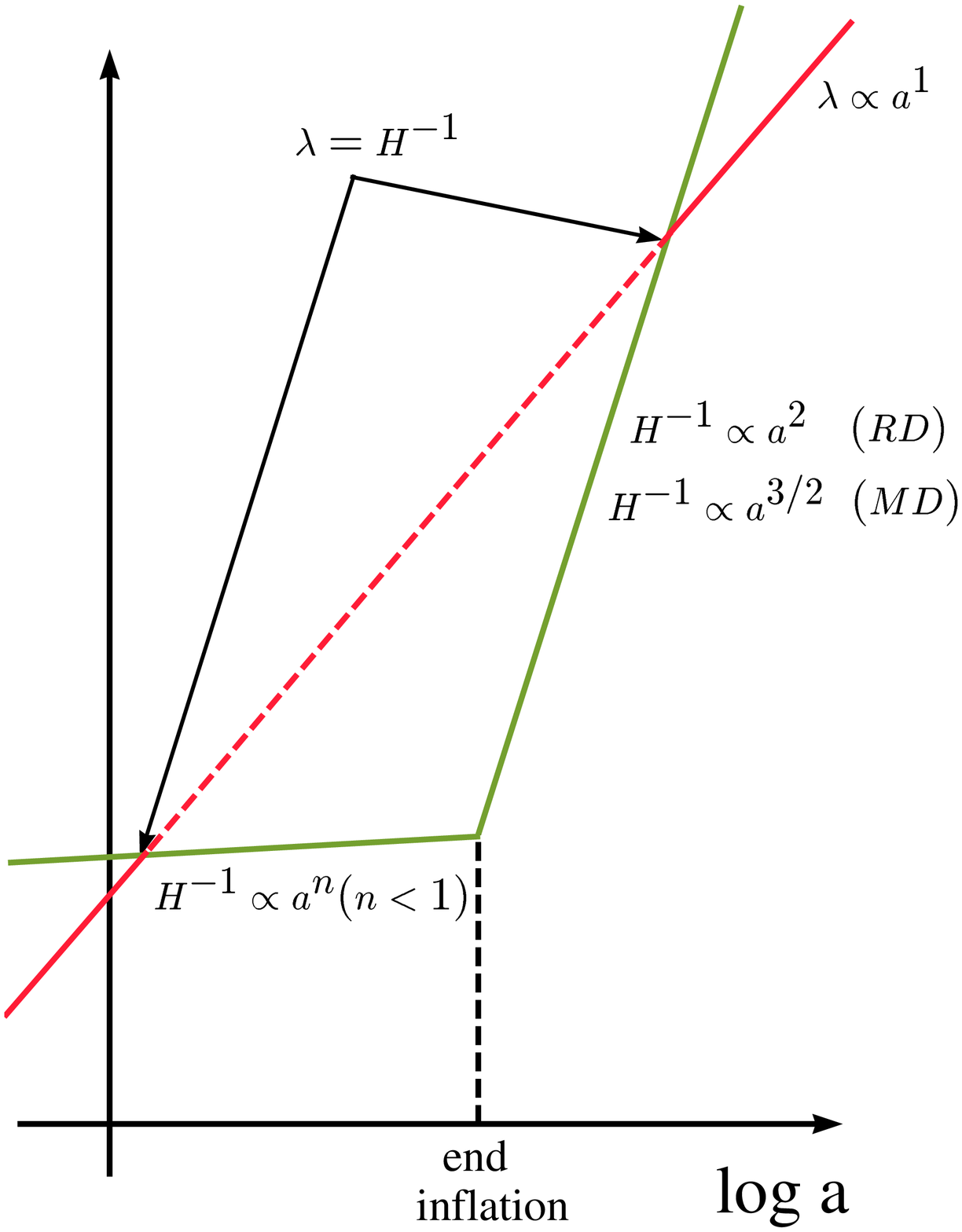}
\caption[fig5]{\label{a_hubble} Physical sizes increase as $a(t)$ in
the expanding universe.  The Hubble radius evolves as
$R_H=H^{-1}=(8\pi G \rho(a) /3)^{1/2}$.  In a radiation-dominated or
matter-dominated universe (illustrated by the left panel) any physical
length scale $\lambda$ starts larger than $R_H$, then crosses the
Hubble radius ($\lambda=H^{-1}$) only once. However, if there was a
period of early inflation when $R_H$ increased more slowly than $a$
(as illustrated in the right panel), it is possible for a physical
length scale to start smaller than $R_H$, become larger than $R_H$,
and after inflation ends become once again smaller than $R_H$.
Periods during which the scale is larger than the Hubble radius are
indicated by the dotted line.}
\end{figure}

On length scales smaller than $R_H$ it is possible to move material
around and make an imprint upon the universe.  Scales larger than
$R_H$ are ``beyond the Hubble radius,'' and the expansion of the
universe prevents the establishment of any perturbation on scales
larger than $R_H$.  

Next consider the evolution of some physical length scale $\lambda$.
Clearly, any physical length scale changes in expansion in proportion
to $a(t)$.

Correlations on physical length scales $\lambda$ larger than $R_H(t)$
are often called {\it acausal}.

Now let us form the dimensionless ratio $L\equiv \lambda/R_H$.  If $L$
is smaller than unity, the length scale is smaller than the Hubble
radius and it is possible to imagine some microphysical process
establishing perturbations on that scale, while if $L$ is larger than
unity, no microphysical process can account for perturbations on that
scale.

Since $R_H=a/\dot{a}$, and $\lambda \propto a$, the ratio $L$ is
proportional to $\dot{a}$, and $\dot{L}$ scales as $\ddot{a}$, which
in turn is proportional to $-(\rho+3p)$.  There are two possible
scenarios for $\dot{L}$ depending upon the sign of $\rho+3p$:
\begin{equation}
\dot{L} \ \left\{  
\begin{array}{lll} 
<0 \rightarrow &  R_H \mbox{ grows faster than } \lambda, 
        & \mbox{happens for }\rho+3p>0 \\
> 0 \rightarrow &  R_H \mbox{ grows more slowly than } \lambda, 
        & \mbox{happens for }\rho+3p  <0.
\end{array}
\right. 
\end{equation}
In the standard MD or RD universe, $\rho+3p>0$, and $R_H$ grows faster
than $\lambda$.

Perturbations that appear to be noncausal at last scattering can be
produced if sometime during the early evolution of the universe the
expansion was such that $\ddot{a}>0$.  If $\ddot{a}>0$, then the
Hubble radius will increase more slowly than any physical scale
increases during inflation.  Physical length scales larger than the
Hubble radius at the time of last scattering would have been smaller
than the Hubble radius during the accelerated era.  Therefore it is
possible to imprint correlations during inflation as a scale passes
out of the horizon and have it appear as an acausal perturbation at
last scattering.

From Einstein's equations, the ``acceleration'' is related to the
energy density and the pressure as $\ddot{a} \propto - (\rho+3p)$.
Therefore, an acceleration (positive $\ddot{a}$) requires an unusual
equation of state with $\rho+3p<0$.  This is the condition for
``accelerated expansion'' or ``inflation.''

We do not yet know when inflation occurred, but the best guess for the
different epochs in the history of the universe is given in Table
\ref{movimenti}.  It is useful to spend a few minutes discussing the
movements of Table \ref{movimenti}.

\renewcommand{\arraystretch}{1.8}

\begin{table}[p]
\begin{center}
\begin{tabular}{||r|c|c|c|c|c|c||}
\hline \hline
tempo & epoch  & age & $\rho$ & $p$ & $\rho+3p$ & relic  \\
\hline \hline
pizzicato & 
\parbox[c]{70pt}{\begin{center}string \\ dominated  \end{center}} &
$\simlt 10^{-43}$s & 
? & ? & ? & ?
\\ \hline
prestissimo &
\parbox[c]{70pt}{\begin{center}vacuum \\ dominated \\ (inflation)\end{center}}
  & $\sim10^{-38}$s & $\rho_V$ & $-\rho_V$ & $-$ & 
\parbox[c]{115pt}{\begin{center}density perturbations \\ gravitational waves \\
		dark matter?\end{center}}
\\ \hline
presto     & \parbox[c]{70pt}{\begin{center}matter \\ dominated  \end{center}}
    &$\sim10^{-36}$s& $\rho_\phi$ & 0 & $+$ &
\parbox[c]{115pt}{\begin{center}phase transitions?\\ dark matter? \\ 
baryogenesis?\end{center}}
\\ \hline 
allegro   & \parbox[c]{70pt}{\begin{center}radiation \\ dominated \end{center}}
     &$\simlt 10^{4}$yr        & $T^4$ & $T^4/3$   &$+$ & 
 \parbox[c]{115pt}{\begin{center} dark matter? \\ baryogenesis? \\ 
        neutrino decoupling  \\ nucleosynthesis \end{center}}
\\ \hline
andante & \parbox[c]{70pt}{\begin{center}matter \\ dominated  \end{center}}
  &$\simgt 10^{4}$yr    & $\rho_M$ & 0               &$+$ &
\parbox[c]{115pt}{\begin{center}recombination \\ radiation decoupling\\ 
growth of structure  \end{center}}
\\ \hline
largo 	& \parbox[c]{70pt}{\begin{center}vacuum \\ dominated \\ (inflation)\\ 
{\em da capo?} \end{center} } & recent	& $\rho_V$ & $-\rho_V$ & $-$ & 
\parbox[c]{75pt}{\begin{center}acceleration of \\ the universe \end{center}} \\
\hline \hline
\end{tabular}
\caption{\label{movimenti} Different epochs in the history of the
universe and the associated tempos of the ever decreasing expansion
rate $H$, along with the equation of state and some of the relics
produced during the various eras.  }
\end{center}
\end{table}

The first movement of the Cosmic Symphony may be dominated by the
string section if on the smallest scales there is a fundamental
stringiness to elementary particles.  If this is true, then the first
movement in the cosmic symphony would have been a pizzicato movement
of vibrating strings about $10^{-43}$s after the bang.  There is
basically nothing known about the stringy phase, if indeed there was
one.  We do not yet know enough about this era to predict relics, or
even the equation of state.

The earliest phase we have information about is the inflationary
phase.  The inflationary movement probably followed the string
movement, lasting approximately $10^{-36}$ seconds.  During inflation
the energy density of the universe was dominated by vacuum energy,
with equation of state $p_V\simeq -\rho_V$.  As we shall see, the best
information we have of the inflationary phase is from the quantum
fluctuations during inflation, which were imprinted upon the metric,
and can be observed as CBR fluctuations and the departures from
homogeneity and isotropy in the matter distribution, {\it e.g.,} the
power spectrum.  Inflation also produces a background of gravitational
radiation, which can be detected by its effect on the CBR, or if
inflation was sufficiently exotic, by direct detection of the relic
background by experiments such as LIGO or LISA.

Inflation was wonderful, but all good things must end.  A lot of
effort has gone into studying the end of inflation (For a review, see
Kofman {\it et al.,} \cite{klsreview}.)  It was likely that there was
a brief period during which the energy density of the universe was
dominated by coherent oscillations of the inflaton field.  During
coherent oscillations the inflaton energy density scales as $a^{-3}$
where $a$ is the scale factor, so the expansion rate of the universe
decreased as in a matter-dominated universe with $p\sim 0$.  Very
little is known about this period immediately after inflation, but
there is hope that one day we will discover a relic.  Noteworthy
events that might have occurred during this phase include
baryogenesis, phase transitions, and generation of dark matter.

We do know that the universe was radiation dominated for almost all of
the first 10,000 years.  The best preserved relics of the
radiation-dominated era are the light elements.  The light elements
were produced in the radiation-dominated universe one second to three
minutes after the bang.\footnote{Although I may speak of time after
the bang, I will not address the issue of whether the universe had a
beginning or not, which in the modern context is the question of
whether inflation is eternal.  For the purpose of this discussion,
time zero of the bang can be taken as some time before the end of
inflation in the region of the universe we observe. }  If the baryon
asymmetry is associated with the electroweak transition, then the
asymmetry was generated in the radiation-dominated era.  The radiation
era is also a likely source of dark matter such as WIMPS or axions.
If one day we can detect the $1.9$\,K neutrino background, it would be
a direct relic of the radiation era.  The equation of state during the
radiation era is $p_R = \rho_R/3$.

The earliest picture of the matter-dominated era is the CBR.
Recombination and matter--radiation decoupling occurred while the
universe was matter dominated.  Structure developed from small
primordial seeds during the matter-dominated era.  The pressure is
negligible during the matter-dominated era.

Finally, if recent determinations of the Hubble diagram from
observations of distant high-redshift Type-I supernovae are correctly
interpreted, the expansion of the universe is increasing today
($\ddot{a}>0$).  This would mean that the universe has recently
embarked on another inflationary era, but with the Hubble expansion
rate much less than the rate during the first inflationary era.

\subsection{Simple Dynamics of Inflation: The Inflaton}

In building inflation models it is necessary to find a mechanism by
which a universe dominated by vacuum energy can make a transition from
the inflationary universe to a matter-dominated or radiation-dominated
universe.  There is some unknown dynamics causing the expansion rate
to change with time.  There may be several degrees of freedom involved
in determining the expansion rate during inflation, but the simplest
assumption is that there is only one dynamical degree of freedom
responsible for the evolution of the expansion rate.

If there is a single degree of freedom at work during inflation, then
the evolution from the inflationary phase may be modeled by the action
of a scalar field $\phi$ evolving under the influence of a potential
$V(\phi)$.  Let's imagine the scalar field is displaced from the
minimum of its potential as illustrated in Fig.\ \ref{large_small}.
If the energy density of the universe is dominated by the potential
energy of the scalar field $\phi$, known as the {\it inflaton}, then
$\rho+3p$ will be negative.  The vacuum energy disappears when the
scalar field evolves to its minimum.  The amount of time required for
the scalar field to evolve to its minimum and inflation to end (or
even more useful, the number of e-folds of growth of the scale factor)
can be found by solving the classical field equation for the evolution
of the inflaton field, which is simply given by $\ddot{\phi} + 3H
\dot{\phi} + dV/d\phi = 0$.

\subsection{Quantum fluctuations}

In addition to the classical motion of the inflaton field, during
inflation there are quantum fluctuations.\footnote{Here I will
continue to assume there is only one dynamical degree of freedom.}
Since the total energy density of the universe is dominated by the
inflaton potential energy density, fluctuations in the inflaton field
lead to fluctuations in the energy density.  Because of the rapid
expansion of the universe during inflation, these fluctuations in the
energy density are frozen into super-Hubble-radius-size perturbations.
Later, in the radiation or matter-dominated era they will come within
the Hubble radius as if they were {\it noncausal} perturbations.

The spectrum and amplitude of perturbations depend upon the nature of
the inflaton potential.  Mukhanov \cite{mfb} has developed a very nice
formalism for the calculation of density perturbations.  One starts
with the action for gravity (the Einstein--Hilbert action) plus a
minimally-coupled scalar inflaton field $\phi$:
\begin{equation}
S = -\int d^4\!x \ \sqrt{-g} \left[ \frac{m^2_{Pl}}{16\pi}R - 
\frac{1}{2}g^{\mu\nu}\partial_\mu\phi\partial_\nu\phi  + V(\phi) \right] \  .
\end{equation} 
Here $R$ is the Ricci curvature scalar.  Quantum fluctuations result
in perturbations in the metric tensor and the inflaton field
\begin{equation}
g_{\mu\nu}  \rightarrow  g_{\mu\nu}^{FRW} + \delta  g_{\mu\nu}\ \ ; \qquad
\phi  \rightarrow  \phi_0 + \delta \phi \ , 
\end{equation}
where $ g_{\mu\nu}^{FRW}$ is the Friedmann--Robertson--Walker metric,
and $\phi_0(t) $ is the classical solution for the homogeneous,
isotropic evolution of the inflaton.  The action describing the
dynamics of the small perturbations can be written as
\begin{equation}
\delta_2S = \frac{1}{2}\int d^4\!x \ \left[ \partial_\mu u \partial^\mu u 
+ z^{-1}\frac{d^2z}{d\tau^2}\ u^2    \right]\ ;  \quad z=a\dot{\phi}/H \  ,
\end{equation} 
{\it i.e.,} the action in conformal time $\tau$ ($d\tau^2 = a^2(t)dt^2)$ for
a scalar field in Minkowski space, with mass-squared $
m_u^2=-z^{-1}d^2z/d\tau^2$.  Here, the scalar field $u$ is a
combination of metric fluctuations $\delta g_{\mu\nu}$ and scalar
field fluctuations $\delta \phi$.  This scalar field is related to the
amplitude of the density perturbation.

The simple matter of calculating the perturbation spectrum for a
noninteracting scalar field in Minkowski space will give the amplitude
and spectrum of the density perturbations.  The problem is that the
solution to the field equations depends upon the background field
evolution through the dependence of the mass of the field upon $z$.
Different choices for the inflaton potential $V(\phi)$ results in
different background field evolutions, and hence, different spectra
and amplitudes for the density perturbations.

Before proceeding, now is a useful time to remark that in addition to
scalar density perturbations, there are also fluctuations in the
transverse, traceless component of the spatial part of the metric.
These fluctuations (known as tensor fluctuations) can be thought of as
a background of gravitons.

Although the scalar and tensor spectra depend upon $V(\phi)$, for most
potentials they can be characterized by $Q_{RMS}^{PS}$ (the amplitude
of the scalar and tensor spectra on large length scales added in
quadrature), $n$ (the scalar spectral index describing the best
power-law fit of the primordial scalar spectrum), $r$ (the ratio of
the tensor-to-scalar contribution to $C_2$ in the angular power
spectrum), and $n_T$ ( the tensor spectral index describing the best
power-law fit of the primordial tensor spectrum).  For single-field,
slow-roll inflation models, there is a relationship between $n_T$ and
$r$, so in fact there are only three independent variables.
Furthermore, the amplitude of the fluctuations often depends upon a
free parameter in the potential, and the spectra are normalized by
$Q_{RMS}^{PS}$.  This leads to a characterization of a wide-range of
inflaton potentials in terms of two numbers, $n$ and $r$.

\subsection{Models of inflation}

A quick perusal of the literature will reveal many models of
inflation.  Some of the familiar names to be found include:
old, new, pre-owned,
chaotic, quixotic, ergodic,
exotic, heterotic, autoerotic, 
natural, supernatural, au natural,
power-law, powerless, power-mad,
one-field, two-field, home-field,
modulus, modulo, moduli,
self-reproducing, self-promoting,
hybrid, low-bred, white-bread,
first-order, second-order, new-world order,
pre-big-bang, no-big-bang, post-big-bang,
D-term, F-term, winter-term,
supersymmetric, superstring, superstitious,
extended, hyperextended, overextended,
D-brane, p-brane, No-brain,
dilaton, dilettante, $\ldots$

Probably the first step in sorting out different models is a
classification scheme.  One proposed classification scheme has two
main types.  Type-I inflation models are models based on a single
inflaton field, slowly rolling under the influence of an inflaton
potential $V(\phi)$.  This may seem like a restrictive class, but in
fact many more complicated models can be expressed in terms of an
equivalent Type-I model.  For instance ``extended'' inflation, which
is a Jordan--Brans--Dicke model with an inflaton field and a JBD
scalar field, can be recast as an effective Type-I model. Anything
that is not Type I is denoted as a Type-II model.

\begin{figure}
\centering
\leavevmode\epsfxsize=150pt  \epsfbox{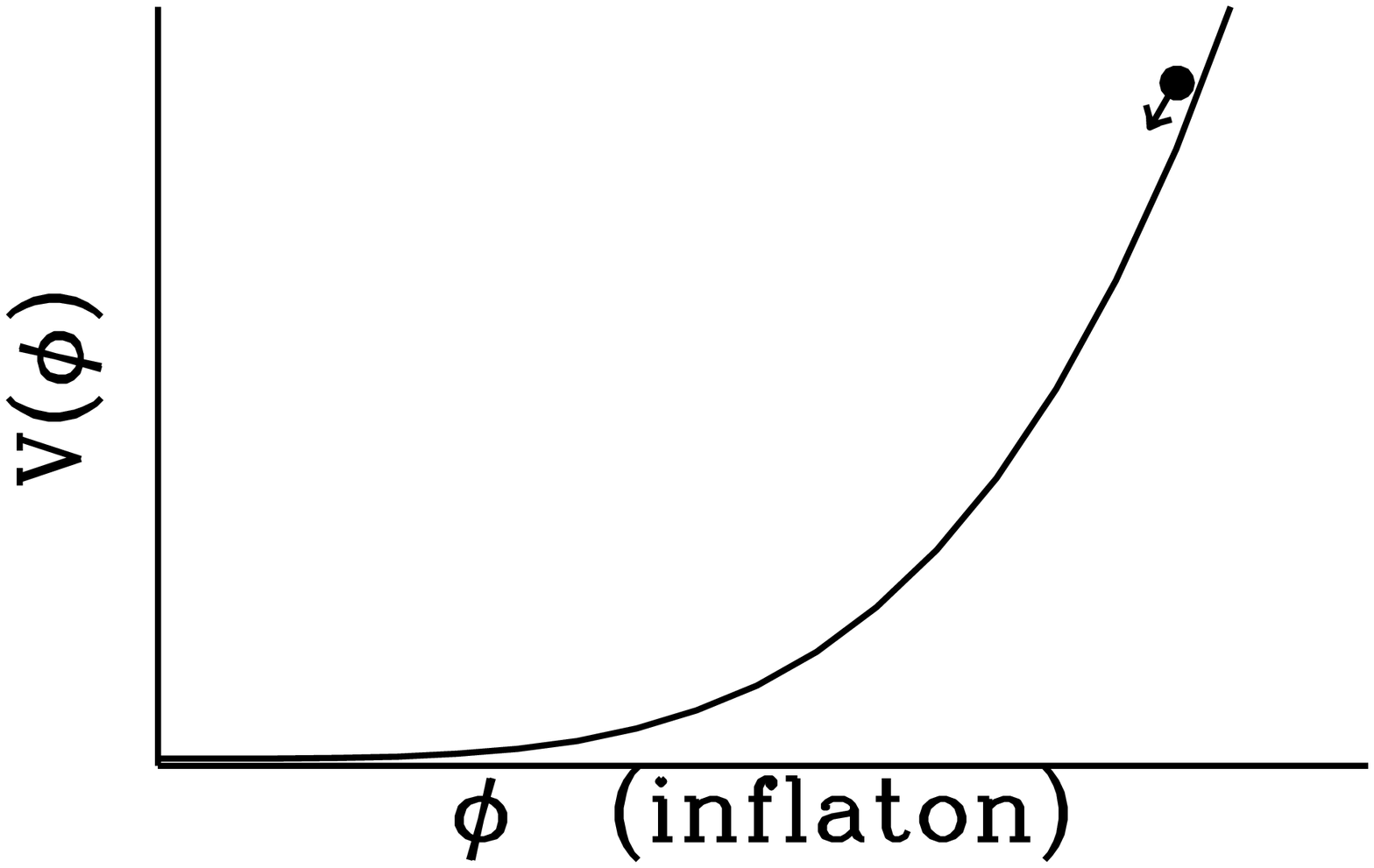}
\leavevmode\epsfxsize=150pt  \epsfbox{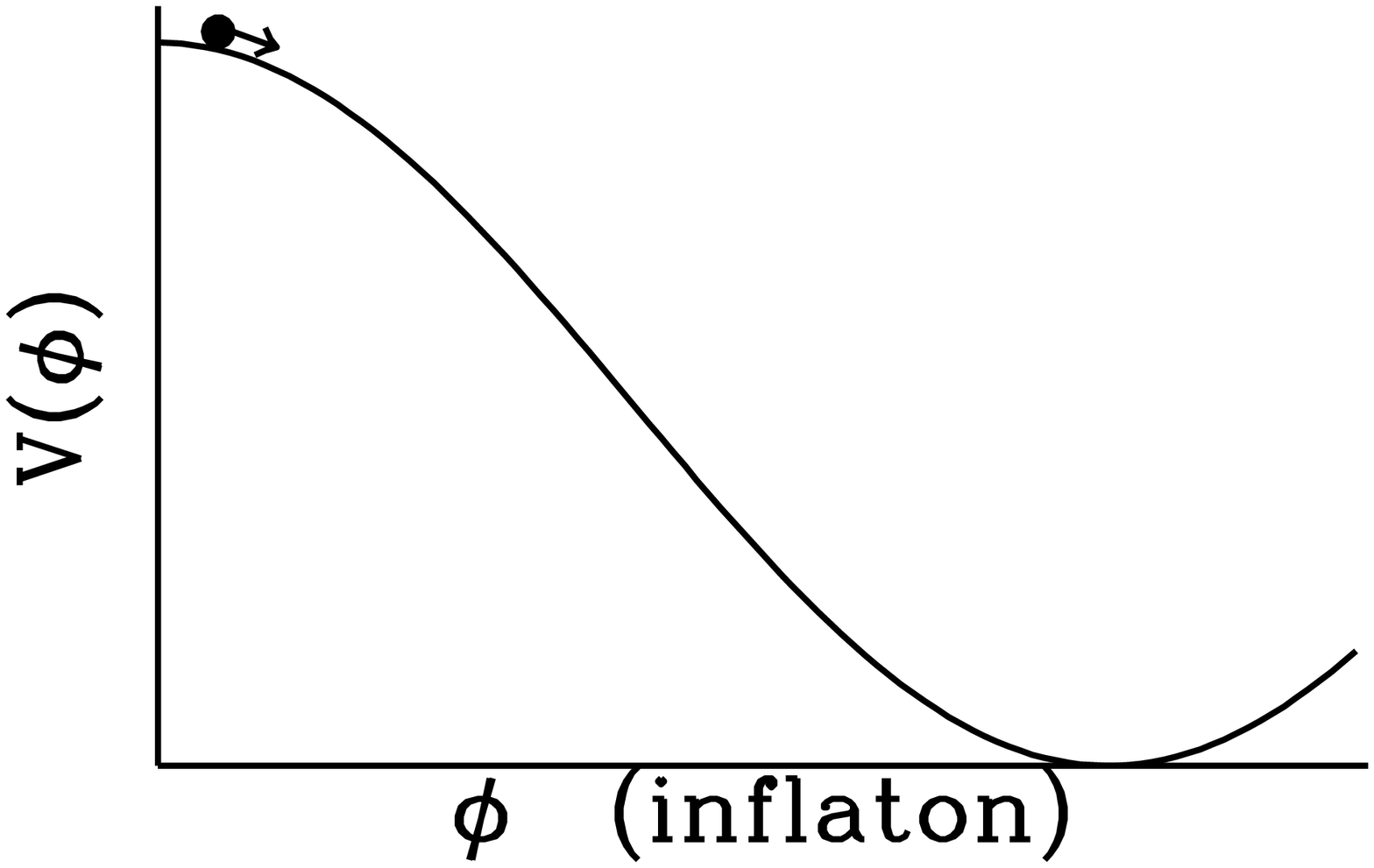} \\
\leavevmode\epsfxsize=150pt  \epsfbox{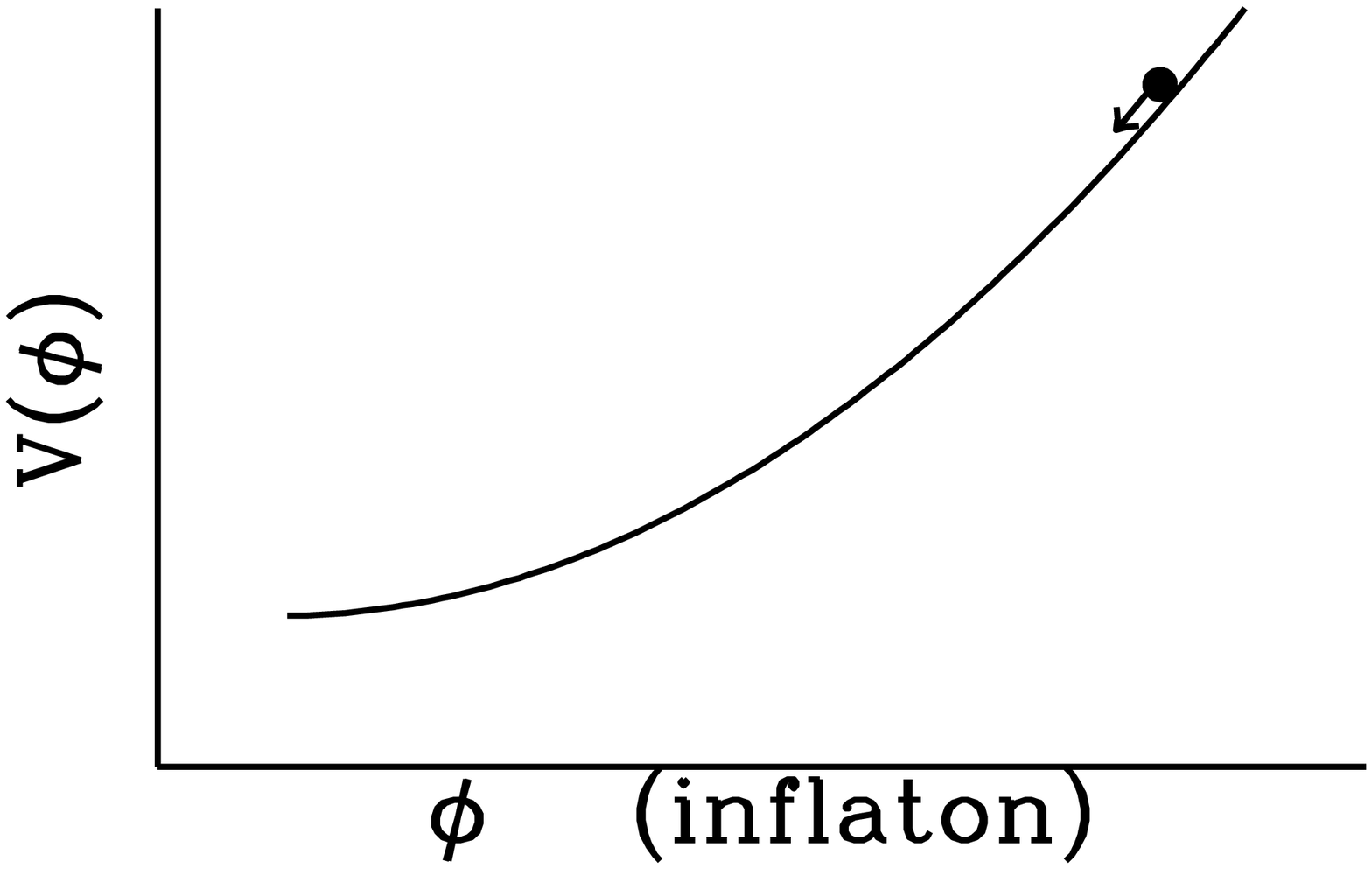}
\caption[fig1]{\label{large_small} Schematic illustrations of the
inflaton potential energy.  The first potential is a ``large-field''
model (Type-Ia).  The second figure illustrates a ``small-field''
model (Type-Ib).  The final figure illustrates an example of hybrid
inflation (Type-Ic).  Notice that the minimum of the potential is
nonzero in the hybrid model.}
\end{figure}

There are subclasses within Type I.  Type-Ia models are
``large-field'' models, where the inflaton field starts large and
evolves toward its minimum.  Examples of large-field models are
chaotic inflation and power-law inflation.  Type-Ib models are
``small-field'' models, where the inflaton field starts small and
evolves to its minimum at larger values of the inflaton field.
Examples of small-field models are new inflation and natural
inflation.  Finally, hybrid-inflation models are classified as Type-Ic
models.  In Type-Ia and Type-Ib models the vacuum energy is
approximately zero at the end of inflation.  Hybrid models have a
significant vacuum energy at the end of inflation (see Fig\
\ref{large_small}).  Hybrid inflation is usually terminated by a
first-order phase transition or by the action of a second scalar
field.  A more accurate description of large-field and small-field
potential is the sign of the second derivative of the potential:
large-field models have $V''>0$ while small-field models have $V''<0$.

\begin{figure}[t]
\centering
\leavevmode\epsfxsize=350pt  \epsfbox{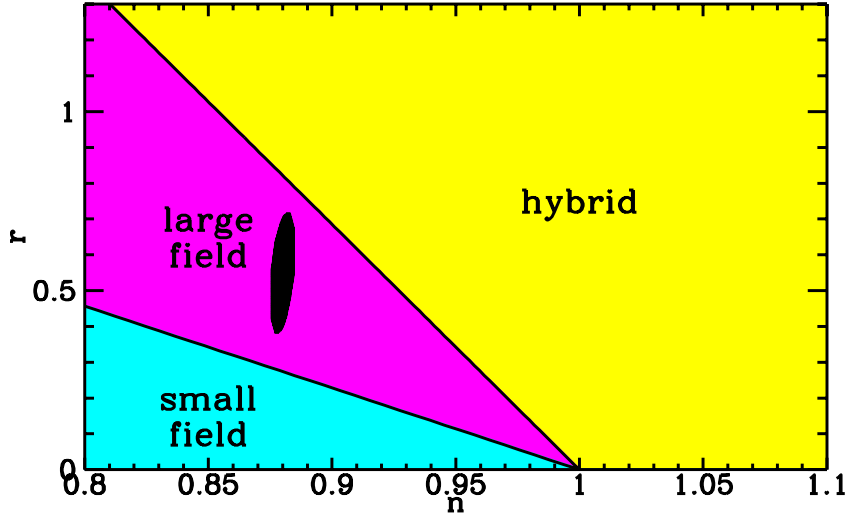}
\caption[fig2]{\label{regions} Large-field (Type-Ia), small-field
(Type-Ib), and hybrid (Type-Ic) models seem to populate different
regions of the $n$--$r$ plane.  Observational determination of $n$ and
$r$ could rule out entire classes of models.  The dark ellipse
indicated the anticipated size of the errors in the post-Planck
era. Of course the location is yet to be determined.  This figure is
from Dodelson, Kinney and Kolb \cite{DKK}.}
\end{figure}

Of course a classification scheme is only reasonable if there are some
observable quantities that can differentiate between different
schemes.  It turns out that the different Type-I models fill in
different regions of the $n$--$r$ plane, as shown in Fig.\
\ref{regions} (from \cite{DKK}).  For a given spectral index $n$,
small-field models have a smaller value of $r$.  Shown as an ellipse
is a very conservative estimate of the uncertainties in $n$ and $r$
that are expected after the next round of satellite observations.
Although we don't know where the error ellipse will fall on the graph,
an error ellipse at the indicated size will restrict models.  So we
expect a true inflation phenomenology, where models of inflation are
confronted by precision observations.

\subsection{Inflation Models in the Era of Precision Cosmology}

It was once said that the words ``precision'' and ``cosmology'' could
not both appear in a sentence containing an even number of negatives.
However that statement is now out of date, or at the very least, very
soon will be out of date.  A number of new instruments will come on
line in the next few years and revolutionize cosmology.

There is now a world-wide campaign to pin down the microwave
anisotropies.  In the near future, long-duration balloon flights, as
well as observations from the cold, dry observatory in Antarctica will
completely change the situation.  Finally, in the next decade two
satellites, a NASA mission ---the Microwave Anisotropy Probe (MAP)---
and an ESA mission---PLANCK, will culminate in a determination of the
spectrum with errors much smaller than the present errors.  Of course
we don't know what the shape of the spectrum will turn out to be, but
we can anticipate errors as small as shown in the figure.

With errors of this magnitude, fitting the spectrum will allow
determination of $n$ and $r$, useful for inflation, as well as
determination of the cosmological parameters ($H_0$, $\Omega_0$,
$\Omega_B$, $\Lambda$, {\it etc.}) to a few percent.

\subsection{Reconstruction}

In addition to restricting the class of inflation models (Type Ia,
Type Ib, {\it etc}.), it may be possible to use the data from
precision microwave experiments to reconstruct a fragment of the
inflationary potential.

Reconstruction of the inflaton potential (see Lidsey, {\it et al.,}
\cite{LLKCBA} for a review) refers to the process of using
observational data, especially microwave background anisotropies, to
determine the inflaton potential capable of generating the
perturbation spectra inferred from observations \cite{Recon}.  Of
course there is no way to prove that the reconstructed inflaton
potential was the agent responsible for generating the perturbations.
What can be hoped for is that one can determine a {\it unique} (within
observational errors) inflaton potential capable of producing the
observed perturbation spectra.  The reconstructed inflaton potential
may well be the first concrete piece of information to be obtained
about physics at scales close to the Planck scale.

As is well known, inflation produces both scalar and tensor
perturbations, and each generate microwave anisotropies (see Liddle
and Lyth \cite{LLrep} for a review).  If $V(\phi)$ is known, the
perturbation spectra can be computed exactly in linear perturbation
theory through integration of the relevant mode equations \cite{GL}.
If the scalar field is rolling sufficiently slowly, the solutions to
the mode equations may be approximated using something known as the
slow-roll expansion \cite{SB,LL,LPB}. The standard reconstruction
program makes use of the slow-roll expansion, taking advantage of a
calculation of the perturbation spectra by Stewart and Lyth \cite{SL},
which gives the next-order correction to the usual lowest-order
slow-roll results.

\begin{table*}
\begin{center}
\begin{tabular}{c|l|l}
\hline \hline 
        & lowest-order      & next-order (exact)   \\
\hline \hline
$V(\phi_*)$     & $H(\phi_*)$
&  $H(\phi_*)$, $\epsilon{(\phi_*)}$  \\
$V'(\phi_*)$    & $H(\phi_*)$, $\epsilon(\phi_*)$ 
& $H(\phi_*)$, $\epsilon(\phi_*)$, $\eta(\phi_*)$ \\
$V''(\phi_*)$   & $H{(\phi_*)}$, $\epsilon(\phi_*)$, $\eta(\phi_*)$
& $H(\phi_*)$, $\epsilon{(\phi_*)}$, $\eta{(\phi_*)}$, $\xi(\phi_*)$  \\
$V'''(\phi_*)$  & $H$, $\epsilon(\phi_*)$, $\eta(\phi_*)$, $\xi(\phi_*)$ 
& \ \ \ \ 
\\
\end{tabular}
\end{center}
\caption{A summary of the slow-roll parameters, $H$, $\epsilon$,
$\eta$, and $\xi$, needed to reconstruct a given derivative of the
potential to a certain order.  Note that the next-order result is
exact. \label{vslowroll}}
\end{table*}

\begin{table*}
\begin{center}
\begin{tabular}{c|l|l}
\hline \hline
parameter & lowest-order & next-order  \\
\hline \hline
$H$ & $A_T^2$                    & $A_T^2$, $A_S^2$    \\
$\epsilon$  &  $A_T^2$, $A_S^2$          & $A_T^2$, $A_S^2$, $n$   \\
$\eta$  & $A_T^2$, $A_S^2$, $n$  & $A_T^2$, $A_S^2$, $n$, $dn/d\ln k$  \\
$\xi$  & $A_T^2$, $A_S^2$, $n$, $dn/d\ln k$ & \ \ \ 
 \\
\end{tabular}
\end{center}
\caption{ The inflation parameters may be expressed in terms of the
primordial scalar and tensor perturbation spectra $A_S^2$, $A_T^2$,
and the scalar and tensor spectral indices $n$, and $dn/d\ln
k$. \label{slowrollobs}}
\end{table*}

Two crucial ingredients for reconstruction are the primordial scalar
and tensor perturbation spectra $A_S(k)$ and $A_T(k)$ which are
defined as in Lidsey {\it et al.,} \cite{LLKCBA}.  The scalar and
tensor perturbations depend upon the behavior of the expansion rate
during inflation, which in turn depends on the value of the scaler
inflaton field $\phi$.  In order to track the change in the expansion
rate, we define slow-roll parameters $\epsilon$, $\eta$, and $\xi$ as
\begin{equation}
\label{epsilon}
\epsilon (\phi) \equiv  \frac{m_{{\rm Pl}}^2}{4\pi} \left[ 
	\frac{H' (\phi) }{H(\phi)} \right]^2 \qquad
\eta (\phi)   \equiv   \frac{m_{{\rm Pl}}^2}{4\pi}  
	\frac{H'' (\phi) }{H(\phi)}  \qquad
\xi(\phi) \equiv \frac{m_{{\rm Pl}}^2}{4\pi} \left( \frac{H'(\phi) H''' 
	(\phi)}{H^2(\phi)} \right)^{1/2} \ ,
\end{equation}
where here the prime superscript implies $d/d\phi$.  If the slow-roll
parameters are small, $\epsilon$, $\eta$, and $\xi$ can be expressed
in terms of derivatives of the inflaton potential:
\begin{equation}
\label{epvd}
\epsilon(\phi)  = 
\frac{m_{{\rm Pl}}^2}{16\pi} \left( \frac{V'}{V} \right)^2 \qquad  
\eta(\phi) = \frac{m_{{\rm Pl}}^2}{8\pi}\, \frac{V''}{V} \qquad
\xi^2 (\phi)  =  
\left( \frac{m^{2}_{{\rm Pl}}}{8\pi}\right)^2\,\frac{V'V^{'''}}{V^{2}}  \ .
\end{equation}

As long as the slow-roll parameters are small compared to unity, the
scalar and tensor perturbation amplitudes $A_S(k)$ and $A_T(k)$ are
given by (see Stewart and Lyth \cite{SL} for the normalization)
\begin{eqnarray}
\label{secondscalar}
A_S^2(k) & \simeq & \frac{4}{25 \pi}\left(\frac{H}{m_{{\rm Pl}}} \right)^2 
\epsilon^{-1} \left[1-(2C+1) \epsilon + C\eta \right]^2  \\
\label{secondtensor}
A_T^2(k) & \simeq & \frac{4}{25 \pi}\left(\frac{H}{m_{{\rm Pl}}} \right)^2
\left[ 1-(C+1 ) \epsilon \right] ^2\ ,
\end{eqnarray}
where $H$, $\epsilon$, and $\eta$ are to be determined at the value of
$\phi$ when $k=aH$ during inflation, and where $C = -2 +\ln 2 + \gamma
\simeq -0.73$ is a numerical constant, $\gamma$ being the Euler
constant.  These equations are the basis of the reconstruction
process.

From the expressions for $A_S^2$ and $A_T^2$ one can express the
scalar and tensor spectral indices, defined as
\begin{equation}
n(k) -1 \equiv 
\frac{d\ln A_S^2}{d\ln k} \qquad n_T(k) = \frac{d\ln A_T^2}{d\ln k} \, 
\end{equation}
in terms of slow-roll parameters.

\begin{table*}
\begin{center}
\begin{tabular}{c|l|l|l}
\hline \hline
        & lowest-order      & next-order &  next-to-next-order   \\
\hline \hline
$V$ & $R_*$, $Q_{RMS}^{PS}$ & $R_*$, $Q_{RMS}^{PS}$ 
	& $R_*$, $Q_{RMS}^{PS}$, $n_*$ \\
$V'$ & $R_*$, $Q_{RMS}^{PS}$ & $R_*$, $Q_{RMS}^{PS}$, $n_*$ & 
             $R_*$, $Q_{RMS}^{PS}$, $n_*$,  $n_*'$ \\
$V''$ & $R_*$, $Q_{RMS}^{PS}$, $n_*$ & $R_*$, $Q_{RMS}^{PS}$, $n_*$,  $n_*'$ 
\\
$V'''$  & $R_*$, $Q_{RMS}^{PS}$, $n_*$, $n_*'$ 
& 
\\
\end{tabular}
\end{center}
\caption{A summary of the observables needed to reconstruct a given
derivative of the potential to a certain order. $R_*$ is the ratio of
the tensor to scalar contributions to the CBR anisotropies at any
conveniently chosen scale.  $Q_{RMS}^{PS}$ is the normalization of the
total (scalar plus tensor) contributions at small $l$, $n_*$ and
$n_*'$ refer to the scalar spectral index and its derivative at some
scale.  Knowledge of $R_*$ and $Q_{RMS}^{PS}$ is equivalent to
determination of $A_T^2(k_*)$ and $A_S^2(k_*)$.  \label{vobs}}
\end{table*}

Perturbative reconstruction requires that one fits an expansion,
usually a Taylor series of the form
\begin{equation}
\ln A_S^2(k)  =  \ln A_S^2(k_*) + (n_*-1) \ln \frac{k}{k_*}   
+ \frac{1}{2}\, \left. \frac{dn}{d\ln k}\right|_* \ln^2\! \frac{k}{k_*}
+ \frac{1}{6} \left. \frac{d^2 n}{d (\ln k)^2} \right|_* \ln^3\! \frac{k}{k_*}
+   \cdots \,,
\end{equation}
where $n_*=n(k_*)$, to the observed spectrum in order to extract the
coefficients, where stars indicate the value at $k_*$. The scale $k_*$
is most wisely chosen to be at the (logarithmic) center of the data,
about $k_* = 0.01 \, h$ Mpc$^{-1}$.

For reconstruction, one takes a Hamilton--Jacobi approach where the
expansion rate is considered fundamental, and the expansion rate is
parameterized by a value of the inflaton field.  The Friedmann
equation may be expressed in terms of the potential $V(\phi)$, the
expansion rate $H(\phi)$, and the derivative of $H$ as
\begin{equation}
 \left( \frac{dH}{d\phi} \right)^2 -\frac{12\pi}{M_{Pl}^2}H^2 
           = -\frac{32\pi^2}{m_{Pl}^4} V(\phi) 
\end{equation}
or equivalently in terms of the slow-roll parameter $\epsilon$, as
\begin{equation}
 V = \frac{m_{Pl}^2H^2}{8\pi}(3-\epsilon) \ .
\end{equation}
Subsequent derivatives of $V$ may be expressed in terms of additional
slow-roll parameters:
\begin{equation}
V'  =  - \frac{m_{Pl}^2}{\sqrt{4\pi}} \, H^2 \epsilon^{1/2}(3-\eta)  \qquad
V''  =  H^2 \left(3\epsilon + 3\eta -\eta^2 + \xi^2 \right) \ ,
\end{equation}
and so on.  So if one can determine the slow-roll parameters, one has
information about the potential.  The slow-roll parameters needed to
reconstruct a given derivative of the potential is given in Table
\ref{vslowroll}.

Of course, the problem is that the slow-roll parameters are not
directly observable!  But one can construct an iterative scheme to
express the slow-roll parameters in terms of observables.  The result
is shown in Table \ref{slowrollobs}.

Using the notation $n'_*$ to indicate $\left.  d n/d\ln k \right|_*$
and $R_*=A_T^2(k_*)/A_S^2(k_*)$, the reconstruction equations are
\begin{eqnarray}
\label{vees}
m_{{\rm Pl}}^{-4}\, V(\phi_*) & \simeq & \frac{75} {32} A_S^2(k_*) R_* 
\left\{ \phantom{\frac{1}{2}} \! \! \! \!1 
+  \left[ 0.21 R_* \right] \, \right\} \nonumber \\ 
m_{{\rm Pl}}^{-3}\, V'(\phi_*) & \simeq & - \frac{75 \sqrt{\pi}}{8} 
A_S^2(k_*) R_*^{3/2}\left\{ 1    \phantom{\frac{1}{2}}  \! \! \! \! - \left[ 
\phantom{R_*^2} \! \! \! \! \! \! \! 0.85R_*  
-  0.53(1-n_*) \right] \,  \right\} \nonumber \\
m_{{\rm Pl}}^{-2}\, V''(\phi_*) & \simeq & \frac{25\pi}{4}  A_S^2(k_*)  R_*
 	\left\{9 R_*  - 1.5(1-n_*)  \phantom{\frac{1}{2}} \! \! \! \! \right. 
   - \left[ 24.3 R_*^2  + 0.25 (1-n_*)^2  \right.
		 \nonumber \\ & & \! \! \! \! \! \! \! \! \! \!
 \left. \phantom{\frac{1}{2}} \left.  \phantom{R_*^2}- 14.8  R_* (1-n_*)  
        -1.6 \, n'_*  \right] \right\}\,   .
\end{eqnarray}
The observables needed to reconstruct a given derivation of the
potential are listed in Table \ref{vobs}.

The biggest hurdle for successful reconstruction is that many
inflation models predict a tensor perturbation amplitude (hence,
$R_*$) well below the expected threshold for detection, and even above
detection threshold the errors can be considerable. If the tensor
modes cannot be identified a unique reconstruction is impossible, as
the scalar perturbations are governed not only by $V(\phi)$, but by
the first derivative of $V(\phi)$ as well.  Knowledge of only the
scalar perturbations leaves an undetermined integration constant in
the non-linear system of reconstruction equations.  Another problem is
that simple potentials usually lead to a nearly exact power-law scalar
spectrum with the spectral index close to unity.  In such a scenario
only a very limited amount of information could be obtained about high
energy physics from astrophysical observations.

However if a tensor mode can be determined, then one may follow the
following reconstruction procedure.  $R_*$, $n_*$, $n'_*$, $\ldots$
are to be determined from observations.  Fortunately, parameter
estimation from the microwave background has been explored in some
detail \cite{parest,ZSS}. We shall use error estimates for Planck
assuming polarized detectors are available, following the analysis of
Zaldarriaga {\it et al.,} \cite{ZSS}.  Most analyses have assumed that
$R_*$ and $n_*$ are the only parameters needed to describe the
spectra.  In a recent paper \cite{CGL}, Copeland, Grivell, and Liddle
have generalized their treatment to allow the power spectrum to
deviate from scale-invariance.  Including extra parameters leads to a
deterioration in the determination of {\em all} the parameters, as it
introduces extra parameter degeneracies.  Fortunately, for most
parameters the uncertainty is not much increased by including the
first few derivatives of $n$ \cite{CGL}, but the parameter $n$ itself
has a greatly increased error bar. If a power-law is assumed it can be
determined to around $\Delta n \simeq 0.004$ \cite{ZSS,CGL}, but
including scale dependence increases this error bar by a factor of ten
or more. Notice that unless one {\em assumes} a perfect power-law
behavior, this increase in uncertainty is applicable even if the
deviation from power-law behavior cannot be detected within the
uncertainty.

{}From Grivell and Liddle \cite{GL}, an estimate of the relevant
uncertainties is
\begin{eqnarray}
\label{obsun}
\Delta (R_*) \simeq 0.004 \; \; & ; & \; \; \Delta (dn/d\ln k) \simeq 0.04 
\nonumber \\
\Delta n \simeq 0.15 \; \; & ; & \; \; \Delta [d^2n/d(\ln k)^2 ]
	\simeq 0.005 \,.
\end{eqnarray}

Once $R_*$, $A_S^2$, $n_*$, and $n'_*$ are determined, then it is
possible to find $V(\phi_*)$, $V'(\phi_*)$, and $V''(\phi_*)$ from
Eq.\ \ref{vees}.  Then one can express $V(\phi)$ as a Taylor series
about $V(\phi_*)$:
\begin{equation}
V(\phi) = V(\phi_*) + V'(\phi_*) \Delta\phi 
+ \frac{1}{2}V''(\phi_*)\Delta\phi^2+\cdots \ .
\end{equation}
$\Delta\phi$ is found from an exact expression connecting changes in
$\phi$ with changes in $k$ \cite{LLKCBA},
\begin{equation}
\label{phi-k}
\frac{d\phi}{d\ln k} = \frac{m_{{\rm Pl}}^2}{4\pi} \frac{H'}{H}
	\frac{1}{\epsilon-1} \simeq \sqrt{\frac{R_*}{4\pi}}\,(1+R_*)\, .
\end{equation}

\begin{figure}[t]
\centering
\leavevmode\epsfxsize=220pt  \epsfbox{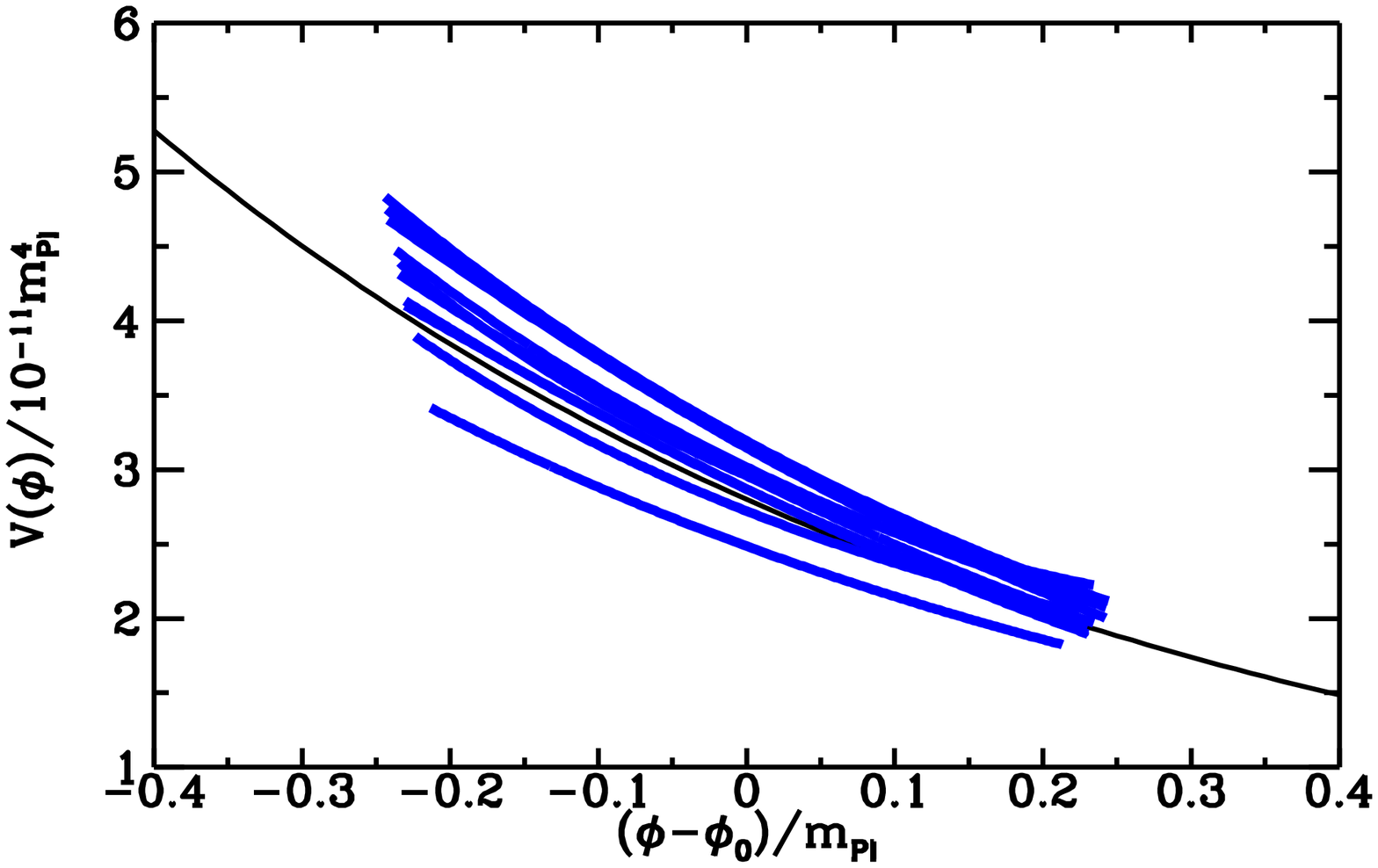} 
\leavevmode\epsfxsize=220pt  \epsfbox{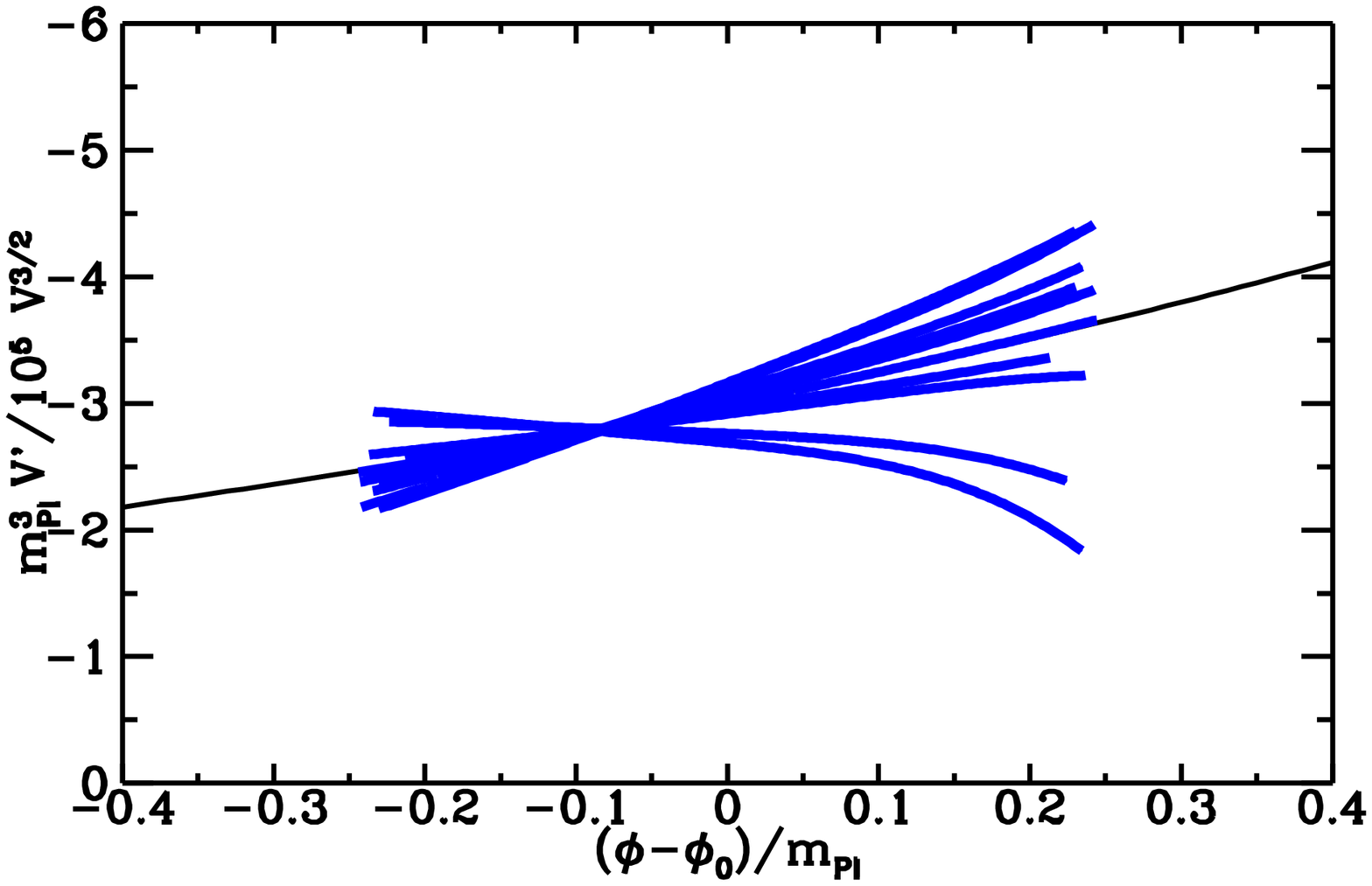} \\
\leavevmode\epsfxsize=220pt  \epsfbox{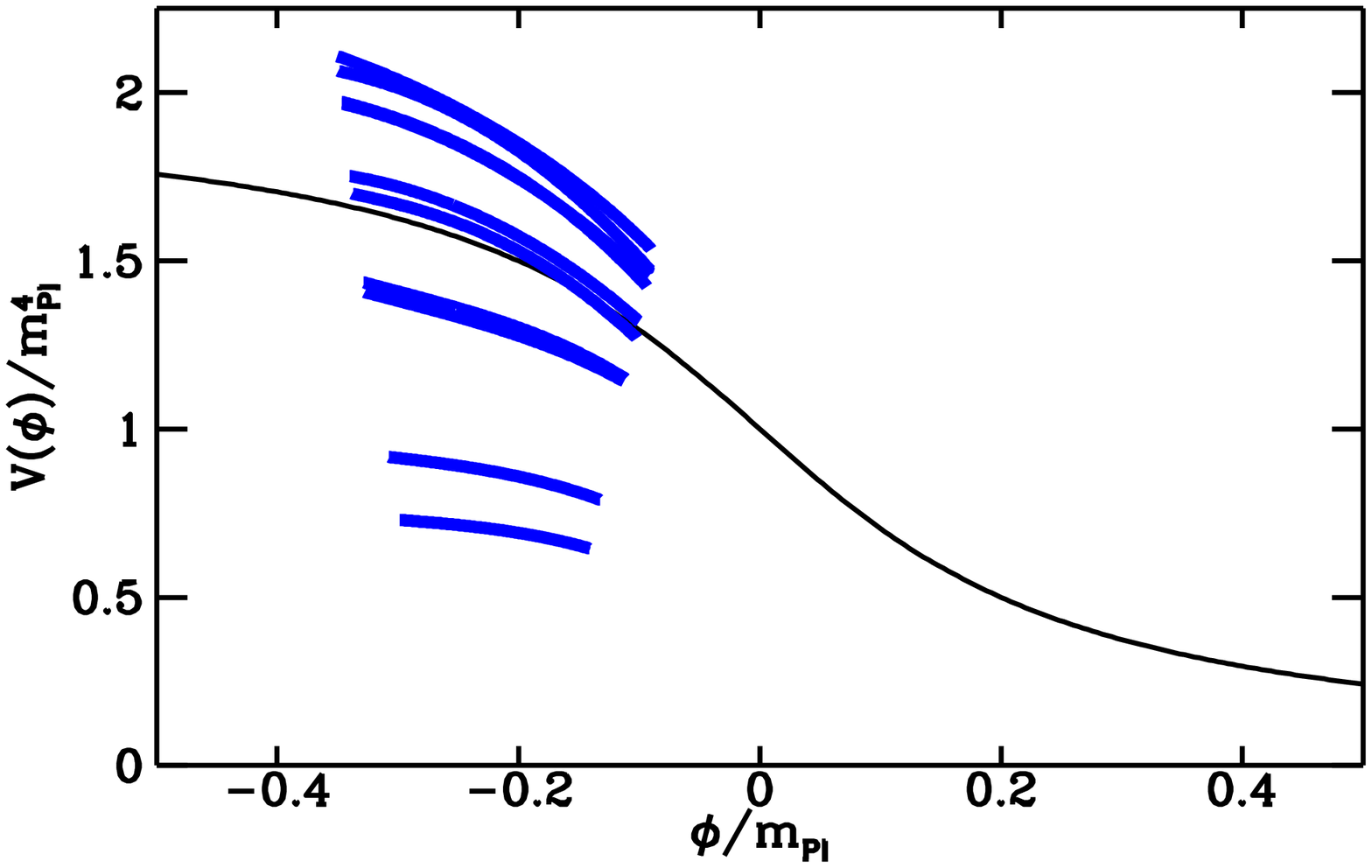} 
\leavevmode\epsfxsize=220pt  \epsfbox{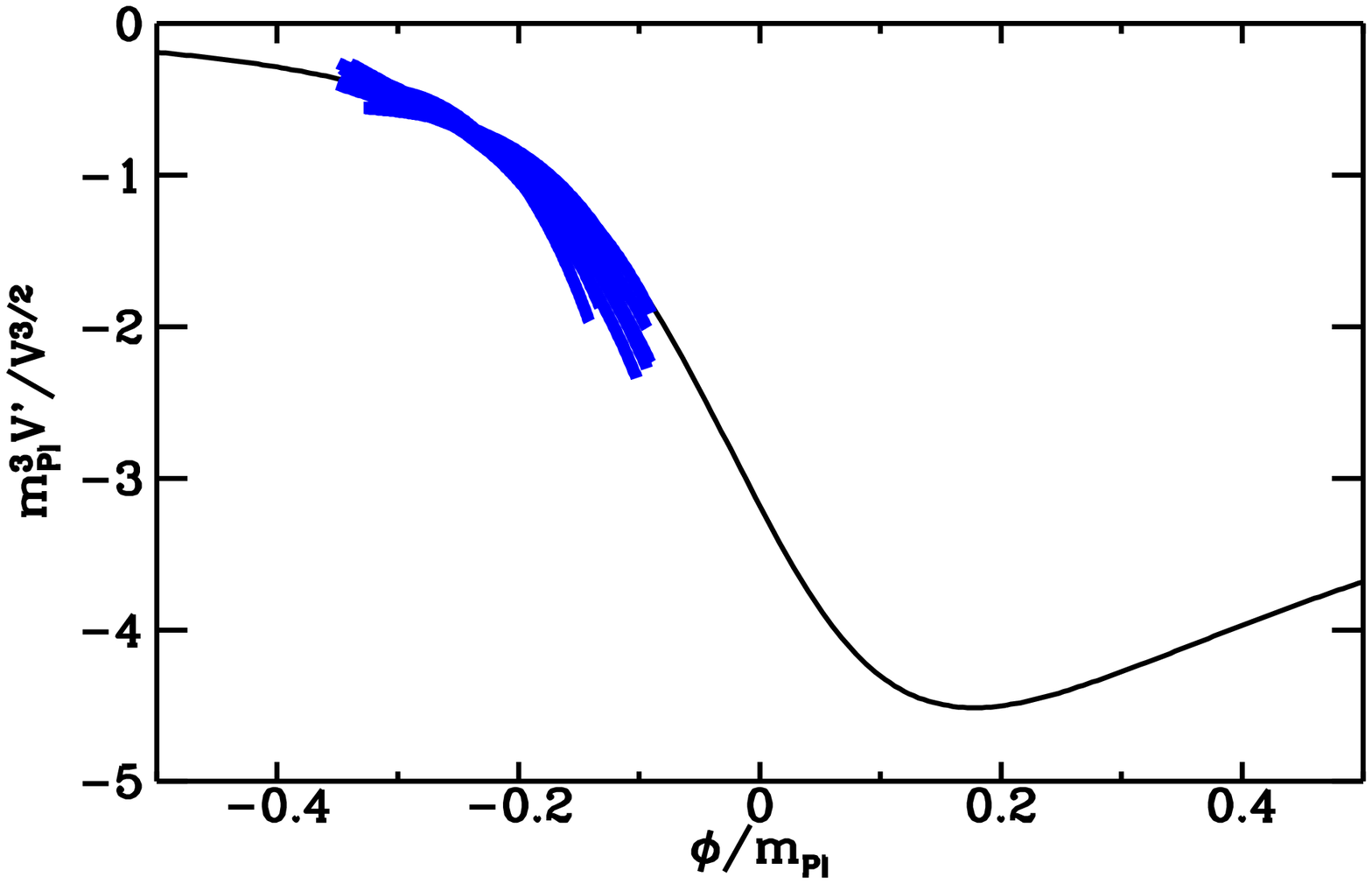} 
\caption[fig3]{\label{reconstruct} Sample reconstructions of two
inflationary potentials.  The upper two figures are the reconstruction
of a power-law potential of the form $V(\phi) =
V_0\exp(-\alpha\phi/m_{Pl})$.  The light curves are from the true
potential, while the heavy curves are ten reconstructions.  The lower
figures are the actual potential and ten reconstructions of a
potential of the form $V(\phi) = \Lambda^4 [1 -(2/\pi) \tan^{-1}
(5\phi/m_{\rm Pl} ) ]$.}
\end{figure}

Let me illustrate reconstruction by considering two sample potentials.
The first potential, discussed by Lidsey {\it et al.,} \cite{LLKCBA}
is a power-law potential, $V(\phi) = V_0\exp(-\alpha\phi/m_{Pl})$,
with $\alpha \simeq 1.6$.  The potential is shown in the upper
left-hand-side of Fig.\ \ref{reconstruct}.  This potential generates a
spectral index of $n_*=0.9$ and $n'_*=0$.  It also results in a value
of $R_*= 0.1$.  So one might guess that precision CBR measurements
will determine $R_*$, $n_*$, and $n'_*$ of these values, with
uncertainties of Eq.\ \ref{obsun}.

In order to see what sort of reconstructed potential results, one can
imagine a model universe with $R_*$, $n_*$, and $n'_*$ generated as
Gaussian random variables with mean determined by the underlying
potential and variance determined by the expected observational
uncertainties.  The result of ten such reconstructions of the
potential are shown in the upper left-hand figure of Fig.\
\ref{reconstruct}.  The range of $\phi$ is determined by the range of
wavenumber over which one expects to have accurate determinations of
the parameters from the CBR.  Here $k_*$ was chosen to be
$0.01h$\,Mpc$^{-1}$ and the range of $k$ taken to be three decades.

Also shown in Fig. \ref{reconstruct} in the upper right-hand panel is
information about the reconstruction of the first derivative of the
potential.

The second example of reconstruction was considered in Copeland {\it
et al.,} \cite{CGKL} The potential, first considered by Wang {\it et
al.,} \cite{WMS} is a potential of the form
\begin{equation}
\label{Vphi}
V(\phi) = \Lambda^4 \left( 1 - \frac{2}{\pi} \, \tan^{-1}
	\frac{5\phi}{m_{{\rm Pl}}} \right) \,,
\end{equation}
which is shown in the lower left-hand side of Fig.\ \ref{reconstruct}. 

A fit to a Taylor expansion to the exact spectrum, using $k_*$ as
above, yields the following results:
\begin{eqnarray}
& & \ln A_S^2(k_*) = 3.84 \quad ; \quad R_*  =  0.014 \quad ; 
 n_*  = 0.579 \quad; \quad \nonumber \\ && \left. \frac{dn}{d \ln k} \right|_* 
=  -0.134
\left. \frac{d^2 n}{d (\ln k)^2} \right|_*  =   -0.052 \ .
\label{params}
\end{eqnarray}
{}From the estimated observational uncertainties of Eq.\
(\ref{obsun}), we see that all these coefficients should be
successfully determined at high significance and a simple
``chi--by--eye'' demonstrates that the spectrum reconstructed from
these data is an adequate fit to the observed spectrum. This stresses
the point that the more unusual a potential is, the more information
one is likely to be able to extract about it, though the uncertainties
on the individual pieces of information may be greater.

We reconstruct the potential in a region about $\phi_* = -0.22$ (the
reconstruction program does not determine $\phi_*$), the width of the
region given via Eq.\ (\ref{phi-k}).  The result for the reconstructed
potential {\em including} the effect of observational errors and
cosmic variance are shown in Fig.\ \ref{reconstruct}, where it can
immediately be seen that the reconstruction has been very successful
in reproducing the main features of the potential while perturbations
on interesting scales are being developed.

The uncertainty is dominated by that of $A_T^2(k_*)$; although the
gravitational waves are detectable in this model, it is only about a
three-sigma detection and the error bar is thus large. Since the
overall magnitude of the potential is proportional to $A_T^2$, the
visual impression is of a large uncertainty.

Fortunately, information in combinations of the higher derivatives is
more accurately determined. Fig.\ \ref{reconstruct} also shows the
reconstruction of $V'/V^{3/2}$ with observational errors; this
combination is chosen as it is independent of the tensors to lowest
order.  Not only is it reconstructed well at the central point, but
both the gradient and curvature are well fit too, confirming useful
information has been obtained about not just $V''$ but $V'''$ as well,
which is only possible because of the extra information contained in
the scale-dependence of the power spectrum. So rather accurate
information is being obtained about the potential.

\section{Preheating, Reheating, and Dark Matter}

If the inflaton is completely decoupled, then once inflation ends it
will oscillate about the minimum of the potential, with the
cycle-average of the energy density decreasing as $a^{-3}$, {\it
i.e.,} as a matter-dominated universe.  But at the end of inflation
the universe is cold and frozen in a low-entropy state: the only
degree of freedom is the zero-momentum mode of the inflaton field.  It
is necessary to ``defrost'' the universe and turn it into a ``hot''
high-entropy universe with many degrees of freedom in the radiation.
Exactly how this is accomplished is still unclear.  It probably
requires the inflaton field to be coupled to other degrees of freedom,
and as it oscillates, its energy is converted to radiation either
through incoherent decay, or through a coherent process involving very
complicated dynamics of coupled oscillators with time-varying masses.
In either case, it is necessary to extract the energy from the
inflaton and convert it into radiation.

I will now turn to a discussion of how defrosting might occur.  It may
be a complicated several-step process.  I will refer to nonlinear
effects in defrosting as ``preheating'' \cite{klsreview} and refer to
linear processes as ``reheating'' \cite{book}.

The possible role of nonlinear dynamics leading to explosive particle
production has recently received a lot of attention.  This process,
known as ``preheating'' \cite{klsreview} may convert a fair fraction
of the inflaton energy density into other degrees of freedom, with
extremely interesting cosmological effects such as symmetry
restoration, baryogenesis, or production of dark matter.  But the
efficiency of preheating is very sensitive to the model and the model
parameters.
 
Perhaps some relic of defrosting, such as symmetry restoration,
baryogenesis, or dark matter may provide a clue of the exact
mechanism, and even shed light on inflation.

\subsection{Defrosting the Universe After Inflation} 

\subsubsection{\label{reheatingsection} Reheating}
In one extreme is the assumption that the vacuum energy of inflation
is immediately converted to radiation resulting in a reheat
temperature $T_{RH}$.

A second (and more plausible) scenario is that reheating is not
instantaneous, but is the result of the slow decay of the inflaton
field.  The simplest way to envision this process is if the comoving
energy density in the zero mode of the inflaton decays into normal
particles, which then scatter and thermalize to form a thermal
background.  It is usually assumed that the decay width of this
process is the same as the decay width of a free inflaton field.

There are two reasons to suspect that the inflaton decay width might
be small.  The requisite flatness of the inflaton potential suggests a
weak coupling of the inflaton field to other fields since the
potential is renormalized by the inflaton coupling to other fields
\cite{review2}.  However, this restriction may be evaded in
supersymmetric theories where the nonrenormalization theorem ensures a
cancellation between fields and their superpartners.  A second reason
to suspect weak coupling is that in local supersymmetric theories
gravitinos are produced during reheating.  Unless reheating is
delayed, gravitinos will be overproduced, leading to a large undesired
entropy production when they decay after big-bang nucleosynthesis
\cite{ellis}.

With the above assumptions, the Boltzmann equations describing the
redshift and interchange in the energy density among the different
components is
\begin{eqnarray}
\label{eq:BOLTZMANNREHEATING}
& &\dot{\rho}_\phi + 3H\rho_\phi +\Gamma_\phi\rho_\phi = 0
	\nonumber \\
& & \dot{\rho}_R + 4H\rho_R - \Gamma_\phi\rho_\phi =0 \ ,
\end{eqnarray}
where dot denotes time derivative.  The dynamics of reheating will be
discussed in detail in Section \ref{reprod}.

The reheat temperature is calculated quite easily \cite{book}.  After
inflation the inflaton field executes coherent oscillations about the
minimum of the potential.  Averaged over several oscillations, the
coherent oscillation energy density redshifts as matter: $\rho_\phi
\propto a^{-3}$, where $a$ is the Robertson--Walker scale factor.  If
we denote as $\rho_I$ and $a_I$ the total inflaton energy density and
the scale factor at the initiation of coherent oscillations, then the
Hubble expansion rate as a function of $a$ is ($M_{Pl}$ is the Planck
mass)
\begin{equation}
H(a) = \sqrt{\frac{8\pi}{3}\frac{\rho_I}{M^2_{Pl}}
	\left( \frac{a_I}{a} \right)^3}\ .
\end{equation}
Equating $H(a)$ and $\Gamma_\phi$ leads to an expression for $a_I/a$.
Now if we assume that all available coherent energy density is
instantaneously converted into radiation at this value of $a_I/a$, we
can define the reheat temperature by setting the coherent energy
density, $\rho_\phi=\rho_I(a_I/a)^3$, equal to the radiation energy
density, $\rho_R=(\pi^2/30)g_*T_{RH}^4$, where $g_*$ is the effective
number of relativistic degrees of freedom at temperature $T_{RH}$.
The result is
\begin{equation}
\label{eq:TRH}
T_{RH} = \left( \frac{90}{8\pi^3g_*} \right)^{1/4}
		\sqrt{ \Gamma_\phi M_{Pl} } \
       = 0.2 \left(\frac{200}{g_*}\right)^{1/4}
	      \sqrt{ \Gamma_\phi M_{Pl} } \ .
\label{eq:trh2}
\end{equation}
The limit from gravitino overproduction is
$T_{RH}\stackrel{<}{{}_\sim} 10^{9}$ to $10^{10}$ GeV.

\subsubsection{\label{preheatsection} Preheating}

The main ingredient of the preheating scenario introduced in the early
1990s is the nonperturbative resonant transfer of energy to particles
induced by the coherently oscillating inflaton fields.  It was
realized that this nonperturbative mechanism can be much more
efficient than the usual perturbative mechanism for certain parameter
ranges of the theory \cite{explosive}.

The basic picture can be seen as follows.  Assume there is an inflaton
field $\phi$ oscillating about the minimum of a potential $V(\phi) =
m^2\phi^2/2$.  It is convenient to parameterize the motion of the
inflaton field as
\begin{equation}
\phi(t) = \Phi(t) \sin mt.
\end{equation}
In an expanding universe, even in the absence if interactions the
amplitude of the oscillations of the inflaton field, $\Phi$, decreases
slowly due to the redshift of the momentum.  In Minkowski space,
$\Phi$ would be a constant in the absence of interactions..

Suppose there is a scalar field $X$ with a coupling to the inflaton of
$ g^2 \phi^2 X^2/2$.  The mode equation for the $X$ field can be
written in terms of a redefined variable $\chi_k \equiv X_k a^{3/2}$
as
\begin{equation}
\ddot{\chi}_k + 3\frac{\dot{a}}{a}\chi_k + \left[ \frac{{\rm \bf k}^2}{a^2} 
+ g^2\Phi^2(t)\sin^2(mt) \right] \chi_k = 0 \ .
\end{equation}

\begin{figure}
\centering
\leavevmode\epsfxsize=350pt  \epsfbox{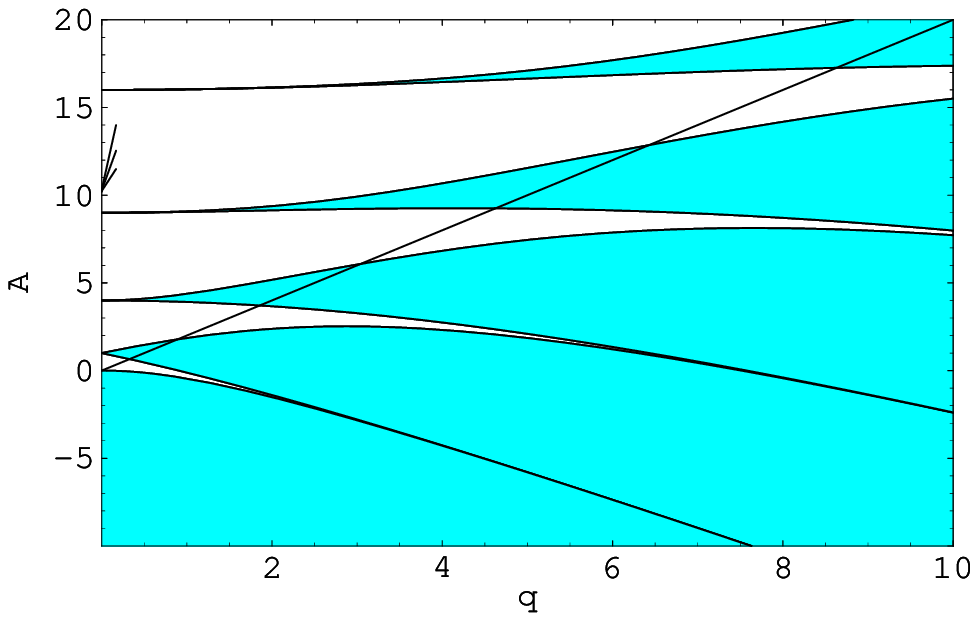}
\caption[figdan]{\label{instability} Shaded areas are regions of
explosive particle production.  The line is $A_k = 2q$ (recall
$A_k=2q+k^2/m^2$).  This figure is from Chung \cite{dansthesis}.}
\vspace*{48pt}
\leavevmode\epsfxsize=350pt  \epsfbox{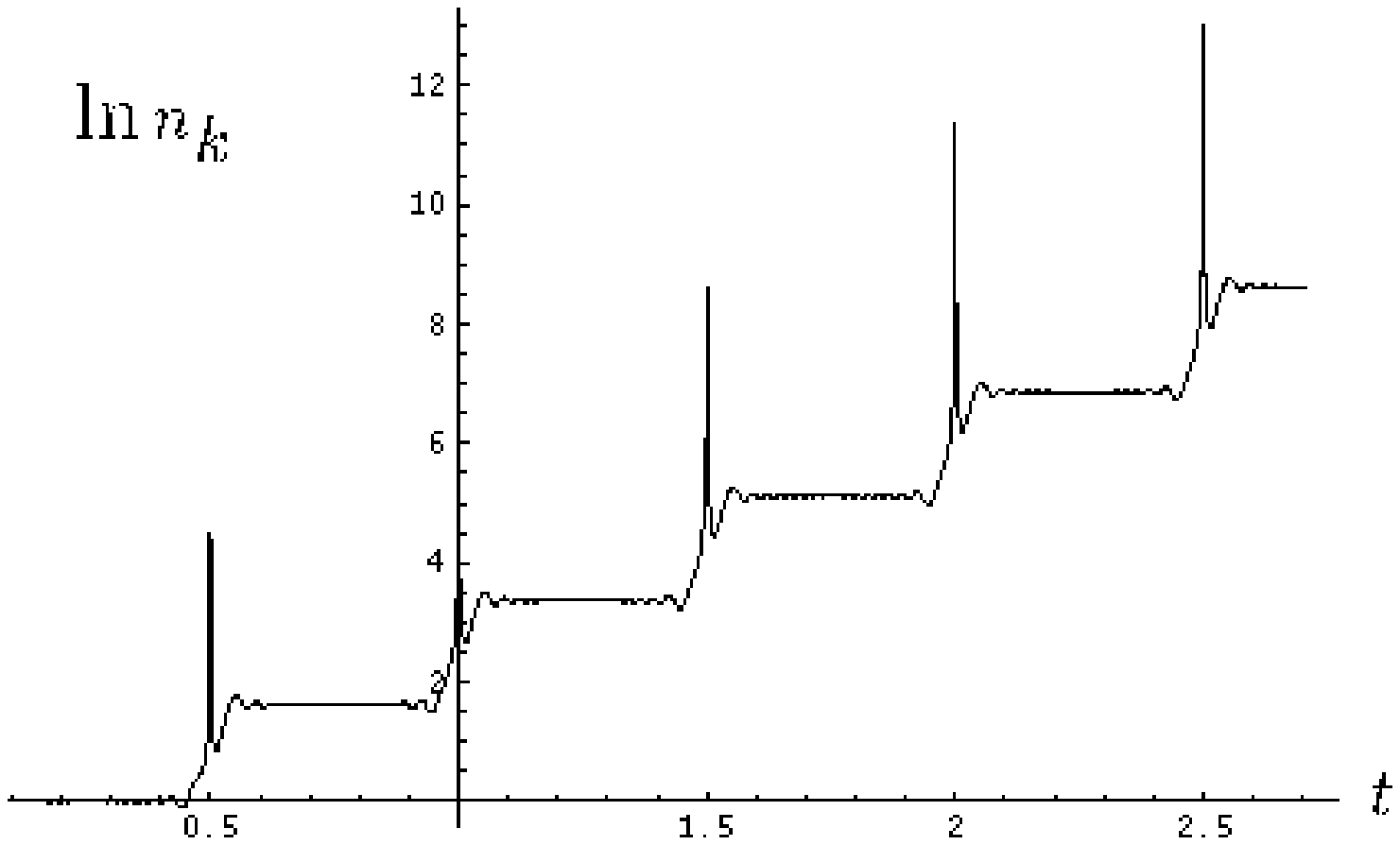}
\caption[kls_1]{\label{kls_1} Explosive particle production if $A_k$
and $q$ are constant.  This figure is from Kofman, Linde, and
Starobinski \cite{klsreview}.}
\end{figure}

In Minkowski space $\dot{a}/a=0$ and $\Phi(t)$ is constant, and the
mode equation becomes (prime denotes $d/dz$ where $z=mt$)
\begin{equation} 
\chi''_k  + [A_k - 2 q \cos(2z)] \chi_k(t)=0 \ .
\label{eq:mathieu}
\end{equation}
The parameter $q$ depends on the inflaton field oscillation amplitude,
and $A_k$ depends on the energy of the particle and $q$:
\begin{eqnarray}
q &= & \frac{g^2\Phi^2}{4m^2} \nonumber \\
A_k & = & 2q + \frac{k^2}{m^2}   \  .
\end{eqnarray}

When $A_k$ and $q$ are constants, the equation is the Mathieu
equation, which exhibits resonant mode instabilities for certain
values of $A_k$ and $q$.  If $A_k$ and $q$ are constant, then there
are instability regions where there is explosive particle production.
The instability regions are shown in Fig.\ \ref{instability} from the
paper of Chung \cite{dansthesis}.

Integration of the field equations for the number of particles created
in a particular $k$ mode is shown in Fig.\ \ref{kls_1}.  Explosive
growth occurs every time the inflaton passes through the origin.

\begin{figure}
\centering
\leavevmode\epsfxsize=350pt  \epsfbox{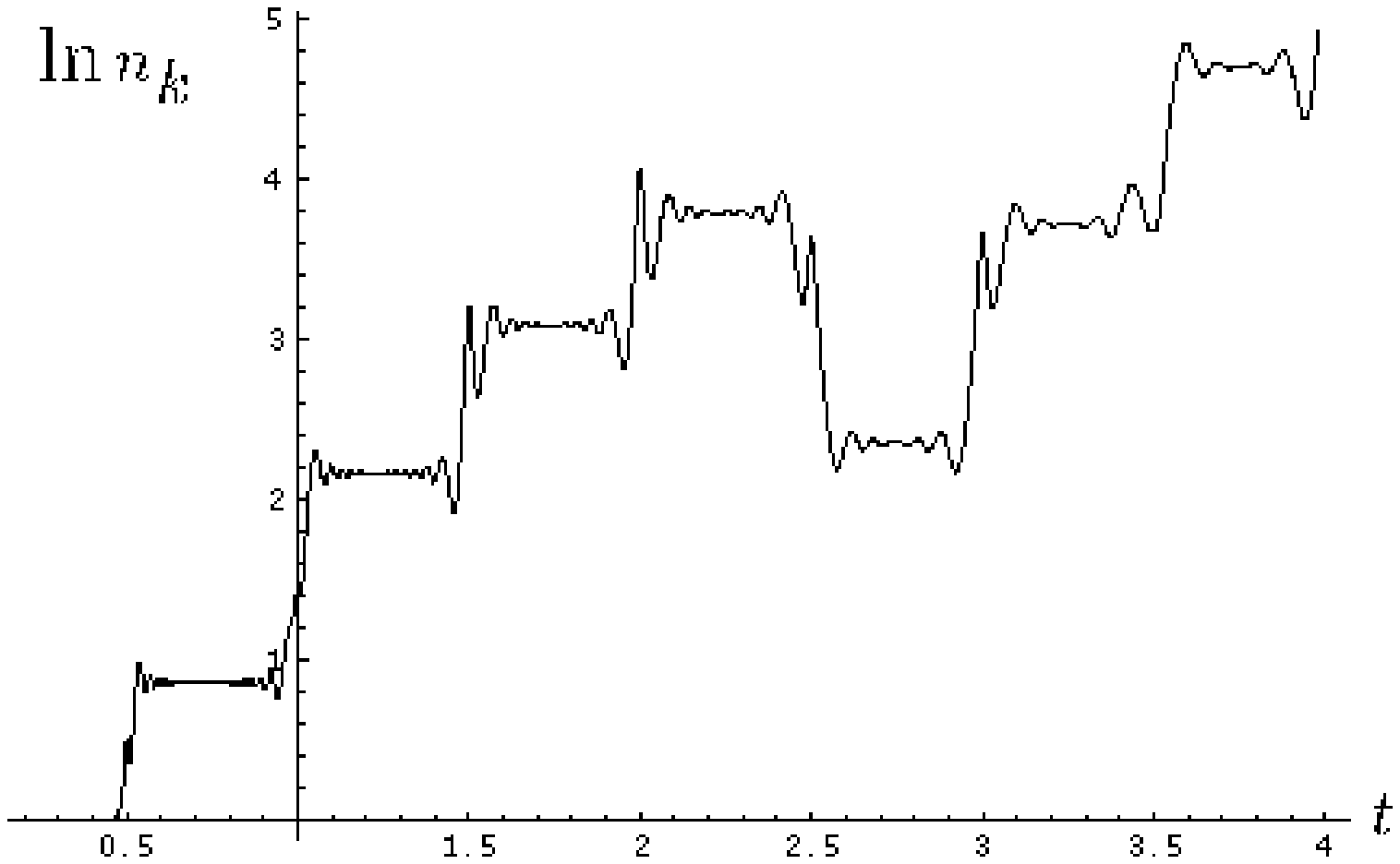}\\
\vspace*{48pt}
\leavevmode\epsfxsize=350pt  \epsfbox{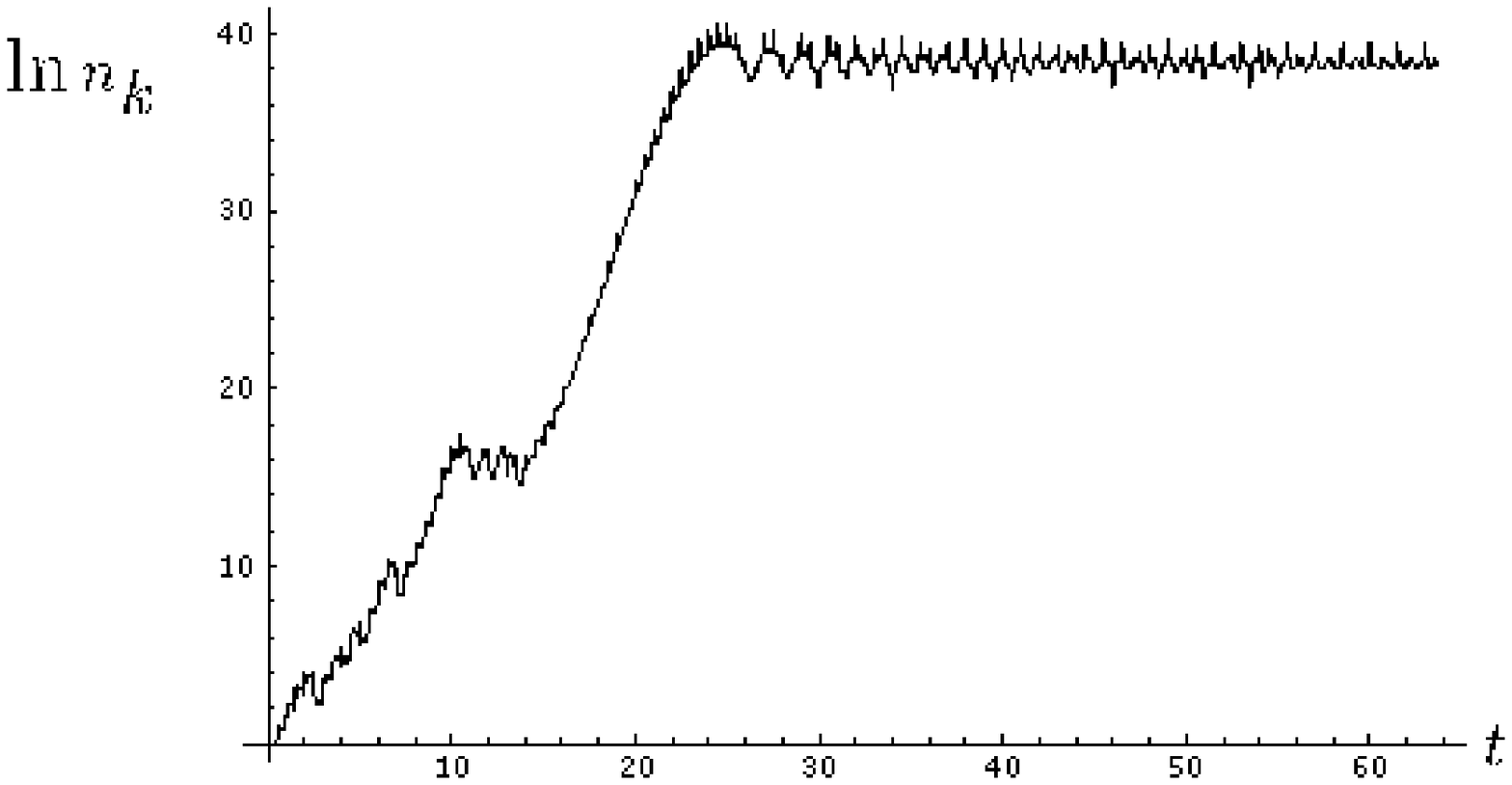}
\caption[kls_2]{\label{kls_2} Explosive particle production if $A_k$
and $q$ are not constant.  The lower figure shows integration for a
longer time.  These figures are from Kofman, Linde, and Starobinski
\cite{klsreview}.}
\end{figure}

In an expanding universe, $A_k$ and $q$ will vary in time, but if they
vary slowly compared to the frequency of oscillations, the effects of
resonance will remain.  An example of particle production where $A_k$
and $q$ vary due to expansion is shown in Fig.\ \ref{kls_2}.  Details
of the parameters and the calculation can be found in Kofman, Linde,
and Starobinski \cite{klsreview}.

If the mode occupation number for the $X$ particles is large, the
number density per mode of the $X$ particles will be proportional to
$|\chi_k|^2$.  If $A_k$ and $q$ have the appropriate values for
resonance, $\chi_k$ will grow exponentially in time, and hence the
number density will attain an exponential enhancement above the usual
perturbative decay.  This period of enhanced rate of energy transfer
has been called preheating primarily because the particles that are
produced during this period have yet to achieve thermal equilibrium.

This resonant amplification leads to an efficient transfer of energy
from the inflaton to other particles which may have stronger coupling
to other particles than the inflaton, thereby speeding up the
reheating process and leading to a higher reheating temperature than
in the usual scenario.

One possible result of preheating is a phase transition caused by the
large number of soft particles created in preheating
\cite{nonthermalphasetransitions}.

Another interesting feature is that particles of mass larger than the
inflaton mass can be produced through this coherent resonant effect.
This has been exploited to construct a baryogenesis scenario
\cite{klr} in which the baryon number violating bosons with masses
larger than the inflaton mass are created through the resonance
mechanism.

\subsection{Dark Matter}
There are many reasons to believe the present mass density of the
universe is dominated by a weakly interacting massive particle
(\WIMP), a fossil relic of the early universe.  Theoretical ideas and
experimental efforts have focused mostly on production and detection
of {\it thermal} relics, with mass typically in the range a few GeV to
a hundred GeV.  Here, I will review scenarios for production of {\it
nonthermal} dark matter.  Since the masses of the nonthermal \WIMPS\
are in the range $10^{12}$ to $10^{19}$ GeV, much larger than the mass
of thermal wimpy \WIMPS, they may be referred to as \WIMPZILLAS.  In
searches for dark matter it may be well to remember that ``size does
matter.''

\subsubsection{Thermal Relics---Wimpy WIMPS}

It is usually assumed that the dark matter consists of a species of a
new, yet undiscovered, massive particle, traditionally denoted by $X$.
It is also often assumed that the dark matter is a thermal relic, {\it
i.e.,} it was in chemical equilibrium in the early universe.

A thermal relic is assumed to be in local thermodynamic equilibrium
(\LTE) at early times.  The {\it equilibrium} abundance of a particle,
say relative to the entropy density, depends upon the ratio of the
mass of the particle to the temperature.  Define the variable $Y\equiv
n_X/s$, where $n_X$ is the number density of WIMP $X$ with mass $M_X$,
and $s \sim T^3$ is the entropy density.  The equilibrium value of
$Y$, $Y_{EQ}$, is proportional to $\exp(-x)$ for $x\gg 1$, while
$Y_{EQ}\sim$ constant for $x\ll 1$, where $x=M_X/T$.

\begin{figure}
\centering
\leavevmode\epsfxsize=300pt  \epsfbox{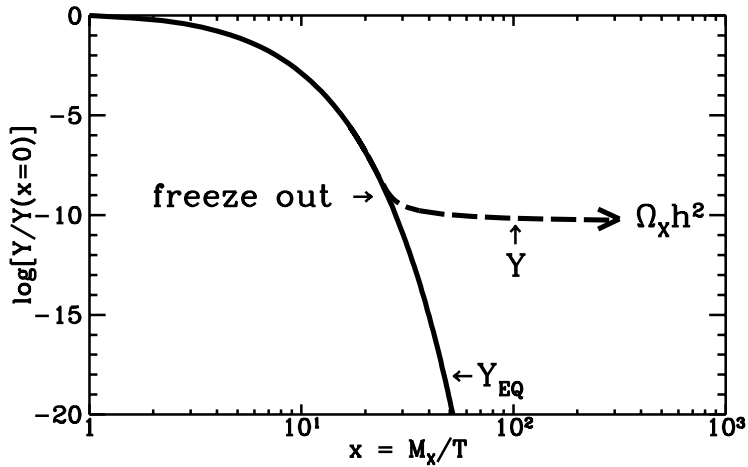}
\caption[fig4]{\label{thermal} A thermal relic starts in \LTE\ at $T\gg
M_X$.  When the rates keeping the relic in chemical equilibrium become
smaller than the expansion rate, the density of the relic relative to
the entropy density freezes out.}
\end{figure}

A particle will track its equilibrium abundance as long as reactions
which keep the particle in chemical equilibrium can proceed rapidly
enough.  Here, rapidly enough means on a timescale more rapid than the
expansion rate of the universe, $H$.  When the reaction rate becomes
smaller than the expansion rate, then the particle can no longer track
its equilibrium value, and thereafter $Y$ is constant.  When this
occurs the particle is said to be ``frozen out.''  A schematic
illustration of this is given in Fig.\ \ref{thermal}.

The more strongly interacting the particle, the longer it stays in
\LTE, and the smaller its eventual freeze-out abundance.  Conversely,
the more weakly interacting the particle, the larger its present
abundance.  The freeze-out value of $Y$ is related to the mass of the
particle and its annihilation cross section (here characterized by
$\sigma_0$) by \cite{book}
\begin{equation}
Y \propto \frac{1}{M_X m_{Pl} \sigma_0} \ .
\end{equation}
Since the contribution to $\Omega$ is proportional to $M_Xn_X$, which
in turn is proportional to $M_XY$, the present contribution to
$\Omega$ from a thermal relic roughly is {\em independent} of its
mass,\footnote{To first approximation the relic dependence depends
upon the mass only indirectly through the dependence of the
annihilation cross section on the mass.}  and depends only upon the
annihilation cross section.  The cross section that results in
$\Omega_Xh^2\sim 1$ is of order $10^{-37}$cm$^2$, i.e., of the order
of the weak scale.  This is one of the attractions of thermal relics.
The scale of the annihilation cross section is related to a known mass
scale.

The simple assumption that dark matter is a thermal relic is
surprisingly restrictive.  The largest possible annihilation cross
section is roughly $M_X^{-2}$.  This implies that large-mass \WIMPS\
would have such a small annihilation cross section that their present
abundance would be too large.  Thus one expects a maximum mass for a
thermal WIMP, which turns out to be a few hundred TeV
\cite{griestkam}.

The standard lore is that the hunt for dark matter should concentrate
on particles with mass of the order of the weak scale and with
interaction with ordinary matter on the scale of the weak force. This
has been the driving force behind the vast effort in dark matter
direct detection.

In view of the unitarity argument, in order to consider {\it thermal}
\WIMPZILLAS, one must invoke, for example, late-time entropy
production to dilute the abundance of these supermassive particles
\cite{k}, rendering the scenario unattractive.

\subsubsection{Nonthermal Relics---WIMPZILLAS}

There are two necessary conditions for the \WIMPZILLA\ scenario.
First, the \WIMPZILLA\ must be stable, or at least have a lifetime
much greater than the age of the universe.  This may result from, for
instance, supersymmetric theories where the breaking of supersymmetry
is communicated to ordinary sparticles via the usual gauge forces
\cite{review}. In particular, the secluded and the messenger sectors
often have accidental symmetries analogous to baryon number. This
means that the lightest particle in those sectors might be stable and
very massive if supersymmetry is broken at a large scale \cite{raby}.
Other natural candidates arise in theories with discrete gauge
symmetries \cite{discrete} and in string theory and M theory
\cite{john,dimitridark98}.

It is useful here to note that \WIMPZILLA\ decay might be able to
account for ultra-high energy cosmic rays above the
Greisen--Zatzepin--Kuzmin cutoff \cite{vadimvaleri,subircr}.  A wimpy
little thermal relic would be too light to do the job, a \WIMPZILLA\
is needed.

\begin{figure}[t]
\centering
\leavevmode\epsfxsize=225pt  \epsfbox{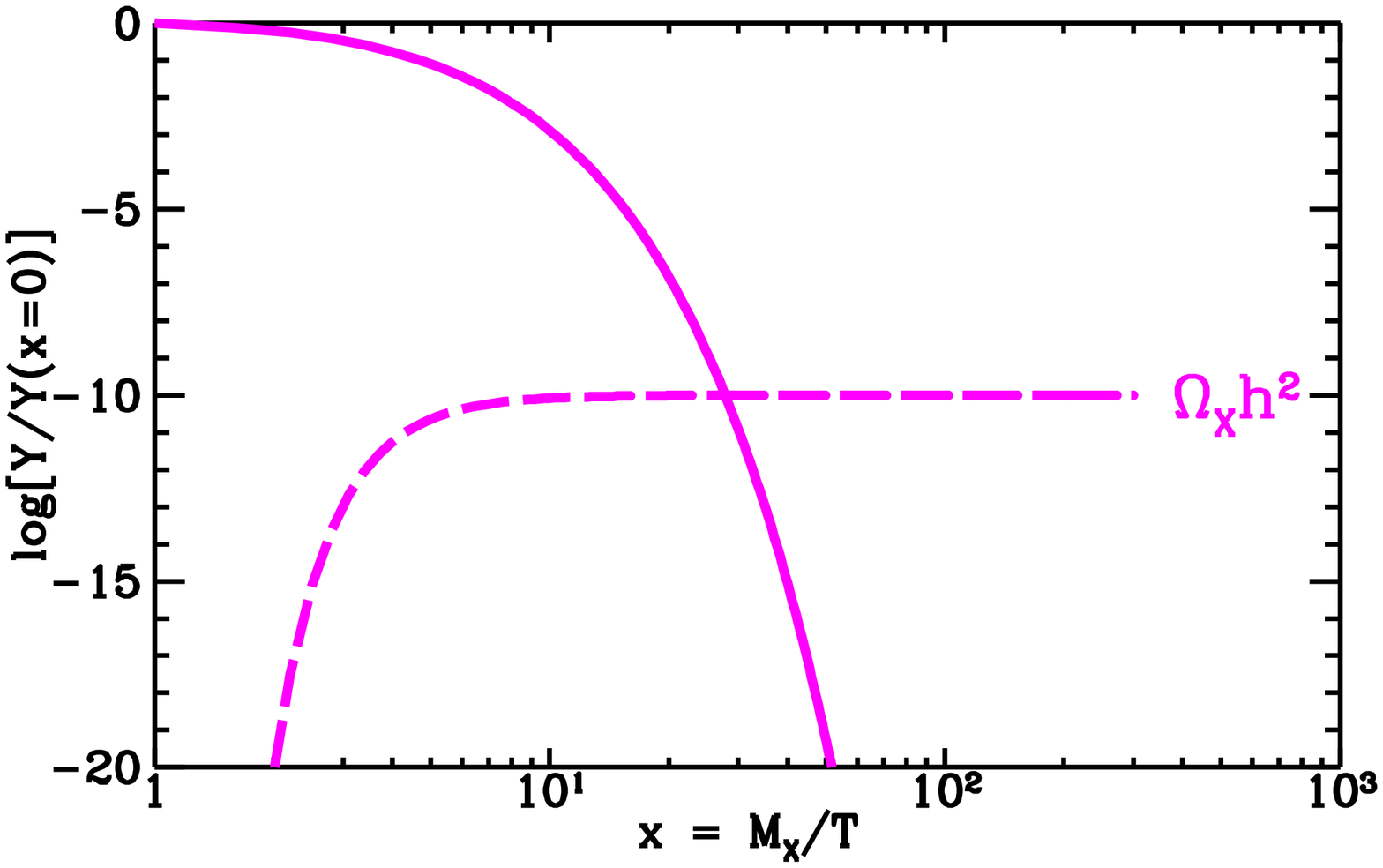}
\leavevmode\epsfxsize=225pt  \epsfbox{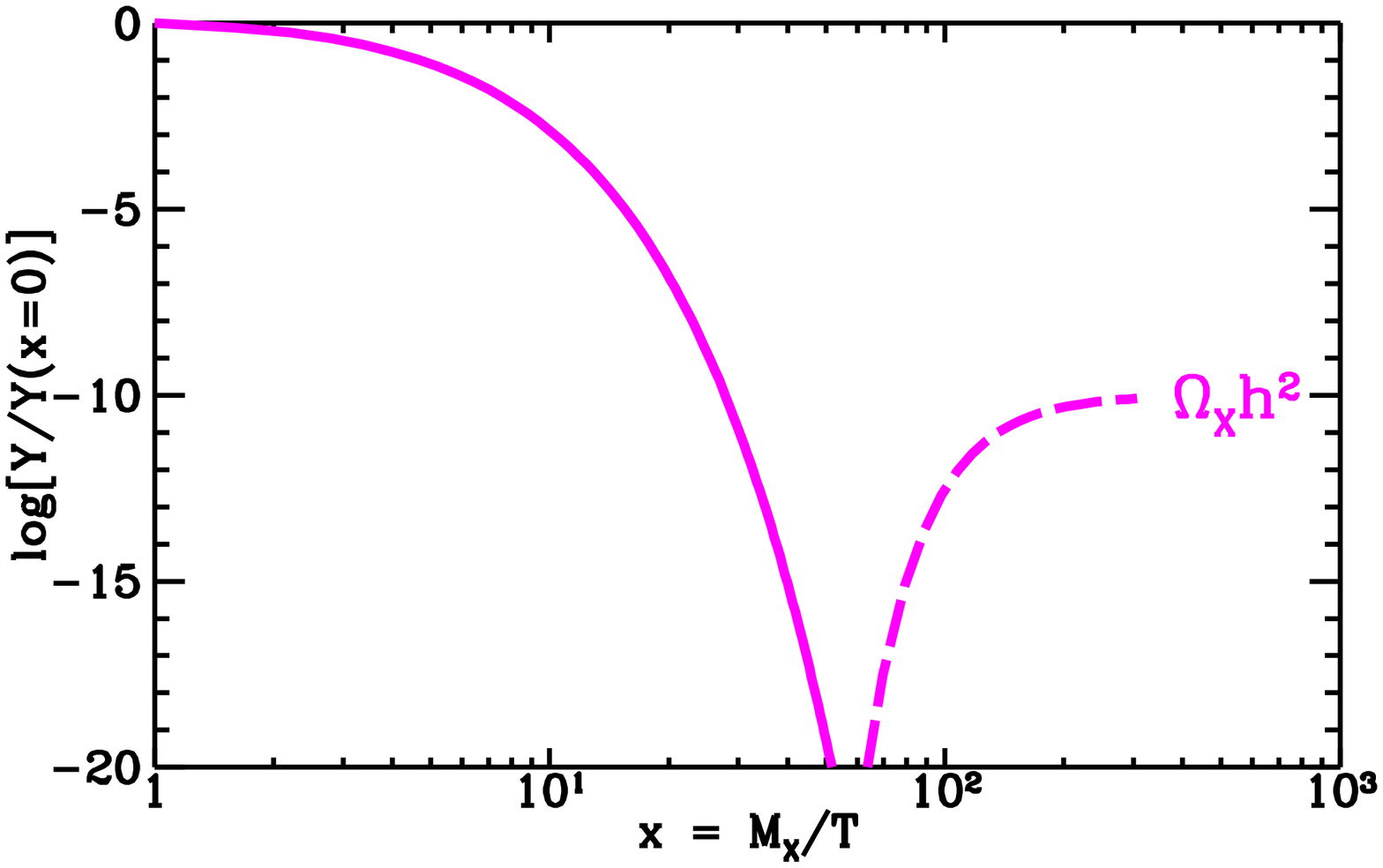}
\caption[fig5]{\label{nonthermal} A nonthermal relic typically has
abundance much less than \LTE\ at $T\gg M_X$.  Here are two examples
of the evolution of a nonthermal relic.  In the left panel $Y\ll
Y_{EQ}$ at freezeout, while in the right panel $Y\gg Y_{EQ}$ at
freezeout. Again, the solid curve is the \LTE\ abundance while the
dashed curve is the actual abundance.}
\end{figure}

The second condition for a \WIMPZILLA\ is that it must not have been in
equilibrium when it froze out ({\it i.e.,} it is not a thermal relic),
otherwise $\Omega_Xh^2$ would be much larger than one.  A sufficient
condition for nonequilibrium is that the annihilation rate (per
particle) must be smaller than the expansion rate: $n_X\sigma|v|<H$,
where $\sigma |v|$ is the annihilation rate times the M{\o}ller flux
factor, and $H$ is the expansion rate.  Conversely, if the dark matter
was created at some temperature $T_*$ {\it and} $\Omega_Xh^2<1$, then
it is easy to show that it could not have attained equilibrium.  To
see this, assume $X$'s were created in a radiation-dominated universe
at temperature $T_*$.  Then $\Omega_Xh^2$ is given by
\begin{equation}
\Omega_Xh^2 = \Omega_\gamma h^2(T_*/T_0)m_Xn_X(T_*)/\rho_\gamma(T_*)\ , 
\end{equation}
where $T_0$ is the present temperature.  Using the fact that
$\rho_\gamma(T_*) = H(T_*) M_{Pl} T_*^2$, $n_X(T_*)/H(T_*) =
(\Omega_X/\Omega_\gamma) T_0 M_{Pl} T_*/M_X$.  One may safely take the
limit $\sigma |v| < M_X^{-2}$, so $n_X(T_*)\sigma |v| / H(T_*)$ must
be less than $(\Omega_X / \Omega_\gamma) T_0 M_{Pl} T_* / M_X^3$.
Thus, the requirement for nonequilibrium is
\begin{equation}
\left( \frac{200\,{\rm TeV}}{M_X}\right)^2 \left( \frac{T_*}{M_X}
\right) < 1 \ .
\end{equation}
This implies that if a nonrelativistic particle with $M_X\simgt 200$
TeV was created at $T_*<M_X$ with a density low enough to result in
$\Omega_X \simlt 1$, then its abundance must have been so small that
it never attained equilibrium. Therefore, if there is some way to
create \WIMPZILLAS\ in the correct abundance to give $\Omega_X\sim 1$,
nonequilibrium is automatic.  Examples of \WIMPZILLA\ evolution and
freezeout are shown in Fig.\ \ref{nonthermal}.

Any \WIMPZILLA\ production scenario must meet these two criteria.
Before turning to several \WIMPZILLA\ production scenarios, it is
useful to estimate the fraction of the total energy density of the
universe in \WIMPZILLAS\ at the time of their production that will
eventually result in $\Omega\sim 1$ today.

The most likely time for \WIMPZILLA\ production is just after
inflation.  The first step in estimating the fraction of the energy
density in \WIMPZILLAS\ is to estimate the total energy density when
the universe is ``reheated'' after inflation.

Consider the calculation of the reheat temperature, denoted as
$T_{RH}$. The reheat temperature is calculated by assuming an
instantaneous conversion of the energy density in the inflaton field
into radiation when the decay width of the inflaton energy,
$\Gamma_\phi$, is equal to $H$, the expansion rate of the universe.
The derivation of the reheat temperature was given in Section
\ref{reheatingsection}.

Now consider the \WIMPZILLA\ density at reheating.  Suppose the
\WIMPZILLA\ never attained \LTE\ and was nonrelativistic at the time
of production.  The usual quantity $\Omega_X h^2$ associated with the
dark matter density today can be related to the dark matter density
when it was produced.  First write
\begin{equation} 
\frac{\rho_X(t_0)}{\rho_R(t_0)}
=\frac{\rho_X(t_{RH})}{\rho_R(t_{RH})}\:\left(\frac{T_{RH}}{T_0}\right),
\label{eq:transfromrh}
\end{equation}
where $\rho_R$ denotes the energy density in radiation, $\rho_X$
denotes the energy density in the dark matter, $T_{RH}$ is the reheat
temperature, $T_0$ is the temperature today, $t_0$ denotes the time
today, and $t_{RH}$ denotes the approximate time of
reheating.\footnote{More specifically, this is approximately the time
at which the universe becomes radiation dominated after inflation.} To
obtain $\rho_X(t_{RH})/\rho_R(t_{RH})$, one must determine when $X$
particles are produced with respect to the completion of reheating and
the effective equation of state between $X$ production and the
completion of reheating.

At the end of inflation the universe may have a brief period of matter
domination resulting either from the coherent oscillations phase of
the inflaton condensate or from the preheating phase
\cite{explosive}.  If the $X$ particles are produced at time
$t=t_{e}$ when the de Sitter phase ends and the coherent oscillation
period just begins, then both the $X$ particle energy density and the
inflaton energy density will redshift at approximately the same rate
until reheating is completed and radiation domination begins.  Hence,
the ratio of energy densities preserved in this way until the time of
radiation domination is
\begin{equation} 
\frac{\rho_X(t_{RH})}{ \rho_R(t_{RH})} \approx \frac{8\pi}{3}\:
\frac{\rho_X(t_{e})}{M_{Pl}^2 H^2(t_{e}) },
\end{equation} 
where $M_{Pl} \approx 10^{19}$ GeV is the Planck mass and most of the
energy density in the universe just before time $t_{RH}$ is presumed
to turn into radiation.  Thus, using Eq.\ \ref{eq:transfromrh}, one
may obtain an expression for the quantity $\Omega_X\equiv
\rho_X(t_0)/\rho_C(t_0)$, where $\rho_C(t_0)=3 H_0^2M_{Pl}^2/8\pi$ and
$H_0=100\: h$ km sec$^{-1}$ Mpc$^{-1}$:
\begin{equation} 
\Omega_X h^2 \approx \Omega_R h^2\:
\left(\frac{T_{RH}}{T_0}\right)\: 
\frac{8 \pi}{3} \left(\frac{M_X}{M_{Pl}}\right)\:
\frac{n_X(t_{e})}{M_{Pl} H^2(t_{e})}.
\label{eq:omegachi}
\end{equation}
Here $\Omega_R h^2 \approx 4.31 \times 10^{-5}$ is the fraction of
critical energy density in radiation today and $n_X$ is the density of
$X$ particles at the time when they were produced.

Note that because the reheating temperature must be much greater than
the temperature today ($T_{RH}/ T_0 \simgt 4.2 \times 10^{14}$), in
order to satisfy the cosmological bound $\Omega_X h^2 \simlt 1$, the
fraction of total \WIMPZILLA\ energy density at the time when they
were produced must be extremely small.  One sees from Eq.\
\ref{eq:omegachi} that $\Omega_Xh^2 \sim 10^{17}(T_{RH}/10^9\mbox
{GeV})(\rho_X(t_e)/\rho(t_e))$.  It is indeed a very small fraction of
the total energy density extracted in \WIMPZILLAS.

This means that if the \WIMPZILLA\ is extremely massive, the challenge
lies in creating very few of them.  Gravitational production discussed
in Section \ref{gravprod} naturally gives the needed suppression.
Note that if reheating occurs abruptly at the end of inflation, then
the matter domination phase may be negligibly short and the radiation
domination phase may follow immediately after the end of inflation.
However, this does not change Eq.\ \ref{eq:omegachi}.

\subsubsection{Wimpzilla Production}

\subsubsubsection{\label{gravprod} Gravitational Production}

First consider the possibility that \WIMPZILLAS\ are produced in the
transition between an inflationary and a matter-dominated (or
radiation-dominated) universe due to the ``nonadiabatic'' expansion of
the background spacetime acting on the vacuum quantum fluctuations
\cite{grav}.

The distinguishing feature of this mechanism is the capability of
generating particles with mass of the order of the inflaton mass
(usually much larger than the reheating temperature) even when the
particles only interact extremely weakly (or not at all) with other
particles and do not couple to the inflaton.  They may still be
produced in sufficient abundance to achieve critical density today due
to the classical gravitational effect on the vacuum state at the end
of inflation.  More specifically, if $0.04 \simlt M_X/H_I \simlt 2$,
where $H_I \sim m_\phi \sim 10^{13}$GeV is the Hubble constant at the
end of inflation ($m_\phi$ is the mass of the inflaton), \WIMPZILLAS\
produced gravitationally can have a density today of the order of the
critical density. This result is quite robust with respect to the
``fine'' details of the transition between the inflationary phase and
the matter-dominated phase, and independent of the coupling of the
\WIMPZILLA\ to any other particle.

Conceptually, gravitational \WIMPZILLA\ production is similar to the
inflationary generation of gravitational perturbations that seed the
formation of large scale structures.  In the usual scenarios, however,
the quantum generation of energy density fluctuations from inflation
is associated with the inflaton field that dominated the mass density
of the universe, and not a generic, sub-dominant scalar field.
Another difference is that the usual density fluctuations become
larger than the Hubble radius, while most of the \WIMPZILLA\
perturbations remain smaller than the Hubble radius.

There are various inequivalent ways of calculating the particle
production due to interaction of a classical gravitational field with
the vacuum (see for example
\cite{fulling,birrelldavies,chitre}). Here, I use the method of
finding the Bogoliubov coefficient for the transformation between
positive frequency modes defined at two different times.  For $M_X/H_I
\simlt 1$ the results are quite insensitive to the differentiability
or the fine details of the time dependence of the scale factor.  For
$0.03 \simlt M_X/H_I \simlt 10$, all the dark matter needed for closure
of the universe can be made gravitationally, quite independently of
the details of the transition between the inflationary phase and the
matter dominated phase.

\begin{figure}
\centering
\leavevmode\epsfxsize=300pt  \epsfbox{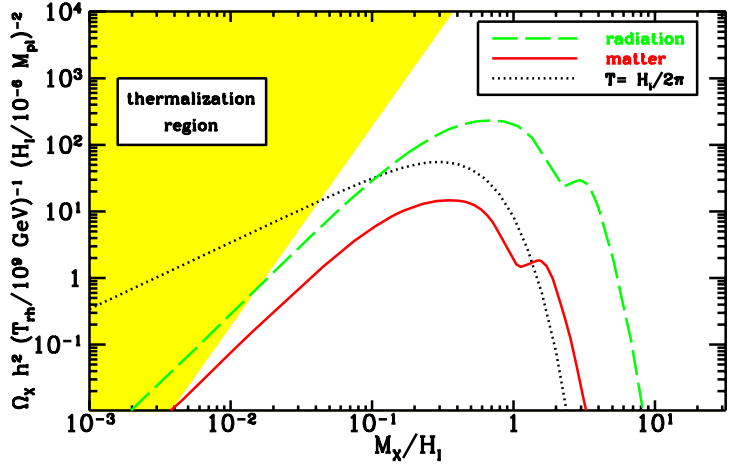}
\caption[fig6]{\label{hawking} The contribution of gravitationally produced
\WIMPZILLAS\ to $\Omega_Xh^2$ as a function of $M_X/H_I$. The shaded
area is where thermalization {\em may} occur if the annihilation cross
section is its maximum value.  Also shown is the contribution assuming
that the \WIMPZILLA\ is present at the end of inflation with a
temperature $T=H_I/2\pi$.}  
\end{figure}

Start with the canonical quantization of the $X$ field in an action of
the form (with metric $ds^2= dt^2- a^2(t) d{\bf x}^2 = a^2(\eta)
\left[d\eta^2 - d{\bf x}^2\right] $ where $\eta$ is conformal time)
\begin{equation}
S=\int dt \int d^3\!x\, \frac{a^3}{2}\left( \dot{X}^2 - \frac{(\nabla
X)^2}{a^2} - M_X^2 X^2 - \xi R X^2 \right)
\end{equation}
where $R$ is the Ricci scalar.  After transforming to conformal time
coordinate, use the mode expansion
\begin{equation}
X({\bf x})=\int \frac{d^3\!k}{(2 \pi)^{3/2} a(\eta)} \left[a_k h_k(\eta) e^{i
{\bf{k \cdot x}}} + a_k^\dagger h_k^*(\eta) e^{-i {\bf{k \cdot x}}}\right],
\end{equation}
where because the creation and annihilation operators obey the
commutator $[a_{k_1}, a_{k_2}^\dagger] = \delta^{(3)}({\bf k}_1 -{\bf
k}_2)$, the $h_k$s obey a normalization condition $h_k h_k^{'*} - h_k'
h_k^* = i$ to satisfy the canonical field commutators (henceforth, all
primes on functions of $\eta$ refer to derivatives with respect to
$\eta$).  The resulting mode equation is
\begin{equation}
h_k''(\eta) + w_k^2(\eta) h_k(\eta) = 0,
\label{eq:modeequation}
\end{equation}
where 
\begin{equation}
w_k^2= k^2 + M_X^2 a^2 + (6 \xi - 1) a''/a \ .
\label{eq:frequency}
\end{equation}
The parameter $\xi$ is 1/6 for conformal coupling and 0 for minimal
coupling.  From now on, $\xi=1/6$ for simplicity but without much loss
of generality.  By a change in variable $\eta \rightarrow k/a$, one
can rewrite the differential equation such that it depends only on
$H(\eta)$, $H'(\eta)/k$, $k/a(\eta)$, and $M_X$.  Hence, the
parameters $H_I$ and $a_I$ correspond to the Hubble parameter and the
scale factor evaluated at an arbitrary conformal time $\eta_I$, which
can be taken to be the approximate time at which $X$s are produced
({\it i.e.,} $\eta_I$ is the conformal time at the end of inflation).

One may then rewrite Eq.\ \ref{eq:modeequation} as
\begin{equation}
h_\ktilde''(\etatilde) + \left(\ktilde^2 + 
\frac{M_X^2}{H_I^2} \tilde{a}^2\right) h_\ktilde(\etatilde) 
=0 \ ,
\label{eq:scaledmodeequation}
\end{equation}
where $\etatilde=\eta a_I H_I$, $\tilde{a}=a/a_I$, and $\ktilde=
k/(a_I H_I)$.  For simplicity of notation,  drop all the
tildes.  This differential equation can be solved once the
boundary conditions are supplied.  

The number density of the \WIMPZILLAS\ is found by a Bogoliubov
transformation from the vacuum mode solution with the boundary
condition at $\eta=\eta_0$ (the initial time at which the vacuum of
the universe is determined) into the one with the boundary condition
at $\eta= \eta_1$ (any later time at which the particles are no longer
being created).  $\eta_0$ will be taken to be $-\infty$ while $\eta_1$
will be taken to be at $+\infty$.  Defining the Bogoliubov
transformation as $ h_k^{\eta_1}(\eta)= \alpha_k h_k^{\eta_0}(\eta) +
\beta_k h_k^{* \eta_0}(\eta)$ (the superscripts denote where the
boundary condition is set), the energy density of produced particles
is
\begin{equation}
 \rho_X(\eta_1) = M_X n_X(\eta_1) = M_X
H_I^3\left (\frac{1}{\tilde{a}(\eta_1)}\right)^3 \int_0^{\infty}
\frac{d\tilde{k}}{2
\pi^2} \tilde{k}^2 |\beta_{\tilde{k}}|^2, 
\end{equation} 
where one should note that the number operator is defined at $\eta_1$
while the quantum state (approximated to be the vacuum state) defined
at $\eta_0$ does not change in time in the Heisenberg representation.

As one can see from Eq.\ \ref{eq:scaledmodeequation}, the input
parameter is $M_X/H_I$.  One must also specify the behavior of
$a(\eta)$ near the end of inflation.  In Fig.\ \ref{hawking} (from
\cite{grav}), I show the resulting values of $\Omega_Xh^2$ as a
function of $M_X/H_I$ assuming the evolution of the scale factor
smoothly interpolates between exponential expansion during inflation
and either a matter-dominated universe or radiation-dominated
universe.  The peak at $M_X/H_I \sim 1$ is similar to the case
presented in Ref.\ \cite{birrelldavies1980}.  As expected, for large
$M_X/H_I$, the number density falls off faster than any inverse power
of $M_X/H_I$.

\begin{figure}
\centering
\leavevmode\epsfxsize=300pt \epsfbox{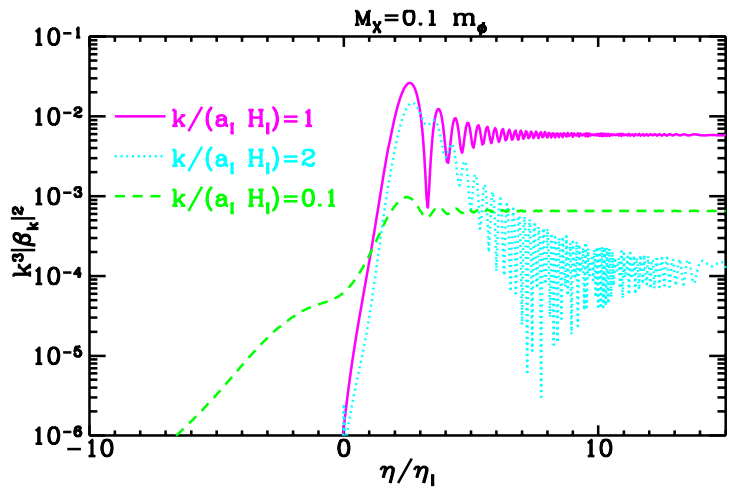}
\caption[fig7]{\label{kofeta} The evolution of the Bogoliubov
coefficient with conformal time for several wavenumbers.
$\eta=\eta_I$ corresponds to the end of the inflationary era.}
\end{figure}

Now most of the action occurs around the transition from inflation to
the matter-dominated or radiation-dominated universe.  This is shown
in Fig.\ \ref{kofeta}.  Also from Fig.\ \ref{kofeta} one can see that
most of the particles are created with wavenumber of order $H_I$.

To conclude, there is a significant mass range ($0.03 H_I$ to $10H_I$,
where $H_I \sim 10^{13}$GeV) for which \WIMPZILLAS\ will have critical
density today regardless of the fine details of the transition out of
inflation.  Because this production mechanism is inherent in the
dynamics between the classical gravitational field and a quantum
field, it needs no fine tuning of field couplings or any coupling to
the inflaton field.  However, only if the particles are stable (or
sufficiently long lived) will these particles give contribution of the
order of critical density.  

\subsubsubsection{\label{reprod} Production During Reheating}

Another attractive origin for \WIMPZILLAS\ is during the defrosting
phase after inflation.  It is important to recall that it is not
necessary to convert a significant fraction of the available energy
into massive particles; in fact, it must be an infinitesimal amount.
I now will discuss how particles of mass much greater than $T_{RH}$
may be created in the correct amount after inflation in reheating
\cite{reheating}.

In one extreme is the assumption that the vacuum energy of inflation
is immediately converted to radiation resulting in a reheat
temperature $T_{RH}$.  In this case $\Omega_X $ can be calculated by
integrating the Boltzmann equation with initial condition $N_X=0$ at
$T=T_{RH}$.  One expects the $X$ density to be suppressed by
$\exp(-2M_X/T_{RH})$; indeed, one finds $\Omega_X \sim 1$ for
$M_X/T_{RH} \sim 25 + 0.5\ln(m_X^2\langle \sigma |v|\rangle)$, in
agreement with previous estimates \cite{vadimvaleri} that for
$T_{RH}\sim10^9$GeV, the \WIMPZILLA\ mass would be about
$2.5\times10^{10}$GeV.

It is simple to calculate the \WIMPZILLA\ abundance in the slow
reheating scenario.  It will be important to keep in mind that what is
commonly called the reheat temperature, $T_{RH}$, is not the maximum
temperature obtained after inflation.  The maximum temperature is, in
fact, much larger than $T_{RH}$.  The reheat temperature is best
regarded as the temperature below which the universe expands as a
radiation-dominated universe, with the scale factor decreasing as
$g_*^{-1/3}T^{-1}$.  In this regard it has a limited meaning
\cite{book,turner}.  One implication of this is that it is incorrect
to assume that the maximum abundance of a massive particle species
produced after inflation is suppressed by a factor of
$\exp(-M/T_{RH})$.

To estimate \WIMPZILLA\ production in reheating, consider a model
universe with three components: inflaton field energy, $\rho_\phi$,
radiation energy density, $\rho_R$, and \WIMPZILLA\ energy density,
$\rho_X$.  Assume that the decay rate of the inflaton field energy
density is $\Gamma_\phi$.  Also assume the \WIMPZILLA\ lifetime is
longer than any timescale in the problem (in fact it must be longer
than the present age of the universe).  Finally, assume that the light
degrees of freedom are in local thermodynamic equilibrium.

With the above assumptions, the Boltzmann equations describing the
redshift and interchange in the energy density among the different
components is [cf.\ Eq.\ (\ref{eq:BOLTZMANNREHEATING})]
\begin{eqnarray}
\label{eq:BOLTZMANN}
& &\dot{\rho}_\phi + 3H\rho_\phi +\Gamma_\phi\rho_\phi = 0
	\nonumber \\
& & \dot{\rho}_R + 4H\rho_R - \Gamma_\phi\rho_\phi
   - 	\frac{\langle\sigma|v|\rangle}{M_X}
	\left[ \rho_X^2 - \left( \rho_X^{EQ} \right)^2 \right] =0
\nonumber \\
& & \dot{\rho}_X + 3H\rho_X 
    + \frac{\langle\sigma|v|\rangle}{M_X}
	\left[ \rho_X^2 - \left( \rho_X^{EQ} \right)^2 \right] = 0  \ ,
\end{eqnarray}
where dot denotes time derivative.  As already mentioned, $\langle
\sigma|v| \rangle$ is the thermal average of the $X$ annihilation
cross section times the M{\o}ller flux factor.  The equilibrium energy
density for the $X$ particles, $\rho_X^{EQ}$, is determined by the
radiation temperature, $T=(30\rho_R/\pi^2g_*)^{1/4}$.

It is useful to introduce two dimensionless constants, $\alpha_\phi$
and $\alpha_X$, defined in terms of $\Gamma_\phi$ and $\langle \sigma
|v| \rangle$ as
\begin{equation}
\label{alphagamma}
\Gamma_\phi = \alpha_\phi M_\phi \qquad
\langle \sigma |v| \rangle = \alpha_X M_X^{-2} \ .
\end{equation}
For a reheat temperature much smaller than $M_\phi$, $\Gamma_\phi$
must be small.  From Eq.\ (\ref{eq:TRH}), the reheat temperature in
terms of $\alpha_X$ and $M_X$ is $T_{RH}\simeq \alpha_\phi^{1/2}
\sqrt{M_\phi M_{Pl}}$.  For $M_\phi=10^{13}$GeV, $\alpha_\phi$ must be
smaller than of order $10^{-13}$.  On the other hand, $\alpha_X$ may
be as large as of order unity, or it may be small also.

It is also convenient to work with dimensionless quantities that can
absorb the effect of the expansion of the universe.  This may be
accomplished with the definitions
\begin{equation}
\label{def}
\Phi \equiv \rho_\phi M_\phi^{-1} a^3 \ ; \quad
R    \equiv \rho_R a^4 \ ; \quad
X    \equiv \rho_X M_X^{-1} a^3 \ .
\end{equation}
It is also convenient to use the scale factor, rather than time, for
the independent variable, so one may define a variable $x = a M_\phi$.
With this choice the system of equations can be written as (prime
denotes $d/dx$)
\begin{eqnarray}
\label{eq:SYS}
\Phi' & = & - c_1 \ \frac{x}{\sqrt{\Phi x + R}}   \ \Phi \nonumber \\
R'    & = &   c_1 \ \frac{x^2}{\sqrt{\Phi x + R}} \ \Phi \
            + c_2 \ \frac{x^{-1}}{ \sqrt{\Phi x +R}} \
	           	         \left( X^2 - X_{EQ}^2 \right) \nonumber \\
X'    & = & - c_3 \ \frac{x^{-2}}{\sqrt{\Phi x +R}} \
		\left( X^2 - X_{EQ}^2 \right) \ .
\end{eqnarray}
The constants $c_1$, $c_2$, and $c_3$ are given by
\begin{equation}
c_1 = \sqrt{\frac{3}{8\pi}} \frac{M_{Pl}}{M_\phi}\alpha_\phi \ \qquad
c_2 = c_1\frac{M_\phi}{M_X}\frac{\alpha_X}{\alpha_\phi} \ \qquad
c_3 = c_2 \frac{M_\phi}{M_X} \ .
\end{equation}
$X_{EQ}$ is the equilibrium value of $X$, given in terms of the
temperature $T$ as (assuming a single degree of freedom for the $X$
species)
\begin{equation}
X_{EQ} = \frac{M_X^3}{M_\phi^3}\left( \frac{1}{2\pi} \right)^{3/2}
	x^3 \left(\frac{T}{M_X}\right)^{3/2}\exp(-M_X/T) \ .
\end{equation}
The temperature depends upon $R$ and $g_*$, the effective number of
degrees of freedom in the radiation:
\begin{equation}
\frac{T}{M_X} = \left( \frac{30}{g_*\pi^2}\right)^{1/4}
\frac{M_\phi}{M_X} \frac{R^{1/4}}{x} \ .
\end{equation}

It is straightforward to solve the system of equations in Eq.\
(\ref{eq:SYS}) with initial conditions at $x=x_I$ of $R(x_I)=X(x_I)=0$
and $\Phi(x_I)=\Phi_I$.  It is convenient to express
$\rho_\phi(x=x_I)$ in terms of the expansion rate at $x_I$, which
leads to
\begin{equation}
\Phi_I = \frac{3}{8\pi} \frac{M^2_{Pl}}{M_\phi^2}
		\frac{H_I^2}{M_\phi^2}\ x_I^3 \ .
\end{equation}
The numerical value of $x_I$ is irrelevant.

Before solving numerically the system of equations, it is useful to
consider the early-time solution for $R$.  Here, early times means $H
\gg \Gamma_\phi$, {\it i.e.,} before a significant fraction of the
comoving coherent energy density is converted to radiation.  At early
times $\Phi \simeq \Phi_I$, and $R\simeq X \simeq 0$, so the equation
for $R'$ becomes $R' = c_1 x^{3/2} \Phi_I^{1/2}$.  Thus, the early
time solution for $R$ is simple to obtain:
\begin{equation}
\label{eq:SMALLTIME}
R \simeq \frac{2}{5} c_1
     \left( x^{5/2} -  x_I^{5/2} \right) \Phi_I^{1/2}
			 \qquad (H \gg \Gamma_\phi) \ .
\end{equation}
Now express $T$ in terms of $R$ to yield the early-time
solution for $T$:
\begin{equation}
\label{threeeights}
\frac{T}{M_\phi}  \simeq  \left(\frac{12}{\pi^2g_*}\right)^{1/4}
c_1^{1/4}\left(\frac{\Phi_I}{x_I^3}\right)^{1/8}  
\left[ \left(\frac{x}{x_I}\right)^{-3/2} -
                \left(\frac{x}{x_I}\right)^{-4} \right]^{1/4}
		\qquad (H \gg \Gamma_\phi) \ .
\end{equation}
Thus, $T$ has a maximum value of
\begin{eqnarray}
\frac{T_{MAX}}{M_\phi}& = & 0.77
   \left(\frac{12}{\pi^2g_*}\right)^{1/4} c_1^{1/4}
   \left(\frac{\Phi_I}{x_I^3}\right)^{1/8} \nonumber \\ & = & 0.77
   \alpha_\phi^{1/4}\left(\frac{9}{2\pi^3g_*}\right)^{1/4} \left(
   \frac{M_{Pl}^2H_I}{M_\phi^3}\right)^{1/4} \ ,
\end{eqnarray}
which is obtained at $x/x_I = (8/3)^{2/5} = 1.48$.  It is also
possible to express $\alpha_\phi$ in terms of $T_{RH}$ and obtain
\begin{equation}
\label{max}
\frac{T_{MAX}}{T_{RH}} = 0.77 \left(\frac{9}{5\pi^3g_*}\right)^{1/8}
		\left(\frac{H_I M_{Pl}}{T_{RH}^2}\right)^{1/4} \ .
\end{equation}

For an illustration, in the simplest model of chaotic inflation $H_I
\sim M_\phi$ with $M_\phi \simeq 10^{13}$GeV, which leads to
$T_{MAX}/T_{RH} \sim 10^3 (200/g_*)^{1/8}$ for $T_{RH} =
10^9$GeV.

\begin{figure}
\centering
\leavevmode\epsfxsize=300pt  \epsfbox{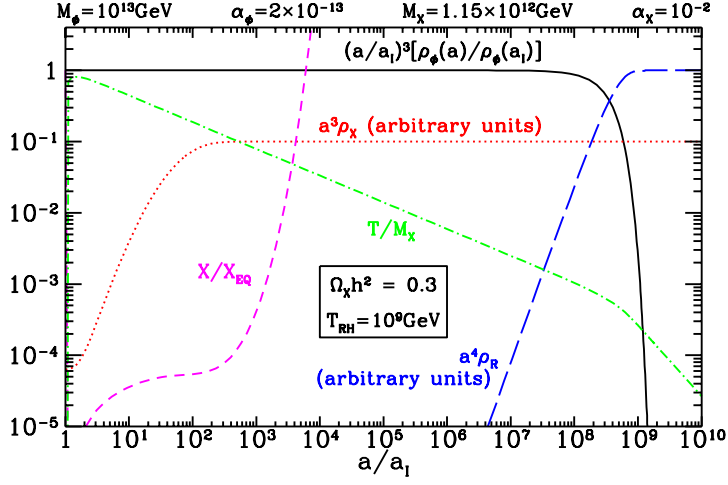}
\caption[fig8]{\label{model1} The evolution of energy densities and $T/M_X$
as a function of the scale factor.  Also shown is $X/X_{EQ}$.}
\end{figure}

We can see from Eq.\ (\ref{eq:SMALLTIME}) that for $x/x_I>1$, in the
early-time regime $T$ scales as $a^{-3/8}$, which implies that entropy
is created in the early-time regime \cite{turner}.  So if one is
producing a massive particle during reheating it is necessary to take
into account the fact that the maximum temperature is greater than
$T_{RH}$, and that during the early-time evolution, $T\propto
a^{-3/8}$.

An example of a numerical evaluation of the complete system in Eq.\
(\ref{eq:SYS}) is shown in Fig.\ \ref{model1} (from \cite{reheating}).
The model parameters chosen were $M_\phi= 10^{13}$GeV, $\alpha_\phi
=2\times10^{-13} $, $M_X= 1.15\times10^{12}$GeV, $\alpha_X =10^{-2}$,
and $g_*=200$.  The expansion rate at the beginning of the coherent
oscillation period was chosen to be $H_I=M_\phi$.  These parameters
result in $T_{RH}=10^9$GeV and $\Omega_Xh^2=0.3$.

Figure \ref{model1} serves to illustrate several aspects of the
problem.  Just as expected, the comoving energy density of $\phi$
({\it i.e.}, $a^3\rho_\phi$) remains roughly constant until
$\Gamma_\phi\simeq H$, which for the chosen model parameters occurs
around $a/a_I\simeq 5\times10^8$.  But of course, that does not mean
that the temperature is zero.  Notice that the temperature peaks well
before ``reheating.''  The maximum temperature, $T_{MAX}= 10^{12}$GeV,
is reached at $a/a_I$ slightly larger than unity (in fact at
$a/a_I=1.48$ as expected), while the reheat temperature, $T_{RH}=
10^9$GeV, occurs much later, around $a/a_I\sim 10^8$.  Note that
$T_{MAX}\simeq 10^3 T_{RH}$ in agreement with Eq.\ (\ref{max}).

{}From the figure it is clear that $X \ll X_{EQ}$ at the epoch of
freeze out of the comoving $X$ number density, which occurs around
$a/a_I\simeq 10^2$.  The rapid rise of the ratio after freeze out is
simply a reflection of the fact that $X$ is constant while $X_{EQ}$
decreases exponentially.

A close examination of the behavior of $T$ shows that after the sharp
initial rise of the temperature, the temperature decreases as
$a^{-3/8}$ [as follows from Eq.\ (\ref{threeeights})] until $H\simeq
\Gamma_\phi$, and thereafter $T\propto a^{-1}$ as expected for the
radiation-dominated era.

For the choices of $M_\phi$, $\alpha_\phi$, $g_*$, and $\alpha_X$ used for
the model illustrated in Fig.\ \ref{model1}, $\Omega_Xh^2 = 0.3$ for  
$M_X=1.15\times10^{12}$GeV, in excellent agreement with the mass predicted 
by using an analytic estimate for the result \cite{reheating}
\begin{equation}
\label{om}
\Omega_X h^2 = M_X^2 \langle \sigma |v|\rangle \,
	\left(\frac{g_*}{200}\right)^{-3/2} \,
	\left (\frac{2000T_{RH}}{M_X}\right)^7 \ .
\end{equation}

Here again, the results have also important implications for the
conjecture that ultra-high cosmic rays, above the
Greisen-Zatsepin-Kuzmin cut-off of the cosmic ray spectrum, may be
produced in decays of superheavy long-living particles
\cite{dimitridark98,vadimvaleri,subircr,kr2}.  In order to produce
cosmic rays of energies larger than about $10^{13}$ GeV, the mass of
the $X$-particles must be very large, $M_X\simgt 10^{13}$ GeV and
their lifetime $\tau_X$ cannot be smaller than the age of the
Universe, $\tau_X\simgt 10^{10}$ yr.  With the smallest value of the
lifetime, the observed flux of ultra-high energy cosmic rays will be
reproduced with a rather low density of $X$-particles, $\Omega_X\sim
10^{-12}$. It has been suggested that $X$-particles can be produced in
the right amount by usual collisions and decay processes taking place
during the reheating stage after inflation if the reheat temperature
never exceeded $M_X$ \cite{kr2}.  Again, assuming naively that that
the maximum number density of a massive particle species $X$ produced
after inflation is suppressed by a factor of $(M_X/T_{RH})^{3/2}
\exp(-M_X/T_{RH})$ with respect to the photon number density, one
concludes that the reheat temperature $T_{RH}$ should be in the range
$10^{11}$ to $10^{15}$GeV \cite{vadimvaleri}. This is a rather high
value and leads to the gravitino problem in generic supersymmetric
models.  This is one reason alternative production mechanisms of these
superheavy $X$-particles have been proposed
\cite{grav,kt,ckr2}. However, our analysis show that the situation is
much more promising. Making use of Eq.\ (\ref{om}), the right amount
of $X$-particles to explain the observed ultra-high energy cosmic rays
is produced for
\begin{equation}
\left(\frac{T_{RH}}{10^{10}\:{\rm GeV}}\right)\simeq
\left(\frac{g_*}{200}\right)^{3/14}\:\left(\frac{M_X}{10^{15}\:{\rm
GeV}}\right),
\end{equation}
where it has been assumed that $\langle \sigma |v|\rangle\sim
M_X^{-2}$. Therefore,  particles as massive as
$10^{15}$ GeV may be generated during the reheating stage in
abundances large enough to explain the ultra-high energy cosmic rays
even if the reheat temperature satisfies the gravitino bound.

\subsubsubsection{\label{preprod} Production During Preheating}

Another way to produce \WIMPZILLAS\ after inflation is in a
preliminary stage of reheating called ``preheating''
\cite{explosive}, where nonlinear quantum effects may lead to an
extremely effective dissipational dynamics and explosive particle
production.

As discussed in Section \ref{preheatsection},
particles can be created in a broad parametric resonance with a
fraction of the energy stored in the form of coherent inflaton
oscillations at the end of inflation released after only a dozen
oscillation periods.  A crucial observation for our discussion is that
particles with mass up to $10^{15}$ GeV may be created during
preheating \cite{kt,klr,gut}, and that their distribution is
nonthermal. If these particles are stable, they may be good candidates
for \WIMPZILLAS\ \cite{dansthesis}.

Interestingly enough, what was found \cite{dansthesis} is that in the
context of a slow-roll inflation with the potential $V(\phi)=m_\phi^2
\phi^2/2$ with the inflaton coupling of $g^2 \phi^2 X^2/2$, the
resonance phenomenon is mostly irrelevant to \WIMPZILLA\ production
because too many particles would be produced if the resonance is
effective.  For the tiny amount of energy conversion needed for
\WIMPZILLA\ production, the coupling $g^2$ must be small enough (for a
fixed $M_X$) such that the motion of the inflaton field at the
transition out of the inflationary phase generates just enough
nonadiabaticity in the mode frequency to produce \WIMPZILLAS.  The
rest of the oscillations, damped by the expansion of the universe,
will not contribute significantly to \WIMPZILLA\ production as in the
resonant case.  In other words, the quasi-periodicity necessary for a
true resonance phenomenon is not present in the case when only an
extremely tiny fraction of the energy density is converted into
\WIMPZILLAS.  Of course, if the energy scales are lowered such that a
fair fraction of the energy density can be converted to \WIMPZILLAS\
without overclosing the universe, this argument may not apply.

The main finding of a detailed treatment \cite{dansthesis} is that
\WIMPZILLAS\ with a mass as large as $10^3 H_I$, where $H_I$ is the
value of the Hubble expansion rate at the end of inflation, can be
produced in sufficient abundance to be cosmologically significant
today.

If the \WIMPZILLA\ is coupled to the inflaton $\phi$ by a term
$g^2\phi^2X^2/2$, then the mode equation in Eq.\ \ref{eq:frequency} is
now changed to
\begin{equation}
\omega_k^2 + k^2 + \left(M_X^2 + g^2\phi^2\right)a^2 \ ,
\end{equation}
again taking $\xi = 1/6$.

The procedure to calculate the \WIMPZILLA\ density is the same
as in Section \ref{gravprod}.  Now, in addition to the parameter
$M_X/H_I$, there is another parameter $g M_{Pl}/H_I$.  Now
in large-field models $H_I\sim 10^{13}$GeV, so $M_{Pl}/H_I$
might be as large as $10^6$.  The choice of $g=10^{-3}$ would
yield $g M_{Pl}/H_I=10^3$.

\begin{figure}
\centering
\leavevmode\epsfxsize=300pt  \epsfbox{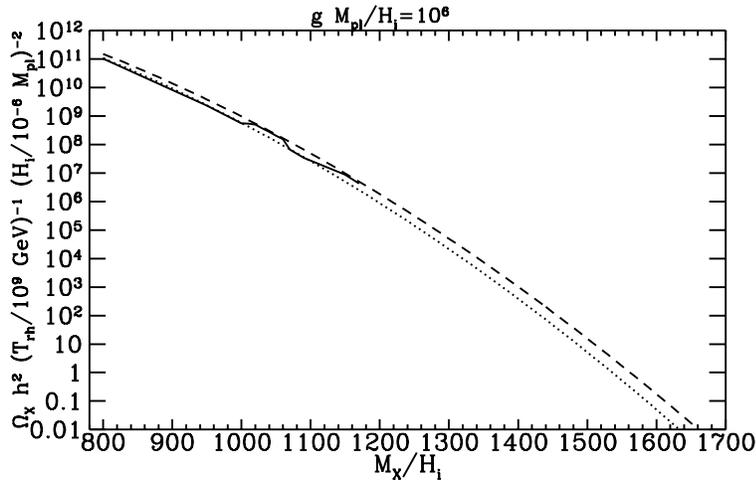} 
\caption[fig9]{\label{dan1}A graph of $\Omega_Xh^2$ versus $M_X/H_I$ for
$gM_{Pl}/H_I= 10^6$.   The solid curve is a numerical result, while
the dashed and dotted curves are analytic approximations \cite{dansthesis}.}
\end{figure}

Fig.\ \ref{dan1} (from \cite{dansthesis}) shows the dependence of the
\WIMPZILLA\ density upon $M_X/H_I$ for the particular choice $g
M_{Pl}/H_I=10^6$.  This would correspond to $g\sim 1$ in large-field
inflation models where $M_{Pl}/H_I=10^6$, about the largest possible
value.  Note that $\Omega_X\sim 1$ obtains for
$M_X/H_I\approx10^3$. The dashed and dotted curves are two analytic
approximations discussed in \cite{dansthesis}, while the solid curve
is the numerical result.  The approximations are in very good
agreement with the numerical results.

\begin{figure}
\centering
\leavevmode\epsfxsize=300pt \epsfbox{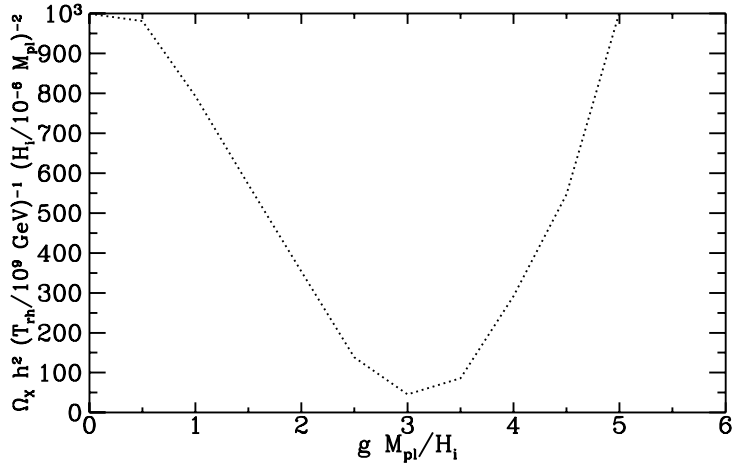}
\caption[fig10]{\label{dan2} An illustration of the nonmonotonic
behavior of the particle density produced with the variation of the
coupling constant.  The value of $M_X/H_I$ is set to unity
\cite{dansthesis}.}
\end{figure}

Fig.\ \ref{dan2} (from \cite{dansthesis}) shows the dependence of the
\WIMPZILLA\ density upon $g M_{Pl}/H_I$.  For this graph $M_X/H_I$ was
chosen to be unity.  This figure illustrates the fact that the
dependence of $\Omega_Xh^2$ on $g M_{Pl}/H_I$ is not monotonic.  For a
detailed explanation of this curious effect, see the paper of Chung
\cite{dansthesis}.

\subsubsubsection{\label{bubprod} Production in Bubble Collisions}

\WIMPZILLAS\ may also be produced \cite{ckr2} if inflation is
completed by a first-order phase transition \cite{ls}, in which the
universe exits from a false-vacuum state by bubble nucleation
\cite{guth}.  When bubbles of true vacuum form, the energy of the
false vacuum is entirely transformed into potential energy in the
bubble walls.  As the bubbles expand, more and more of their energy
becomes kinetic as the walls become highly relativistic.

In bubble collisions the walls oscillate through each other
\cite{moss} and their kinetic energy is dispersed into low-energy
scalar waves \cite{moss,wat}.  We are interested in the potential
energy of the walls, $M_P = 4\pi\eta R^2$, where $\eta$ is the energy
per unit area of a bubble wall of radius $R$.  The bubble walls can be
visualized as a coherent state of inflaton particles, so the typical
energy $E$ of the products of their decays is simply the inverse
thickness of the wall, $E\sim \Delta^{-1}$. If the bubble walls are
highly relativistic when they collide, there is the possibility of
quantum production of nonthermal particles with mass well above the
mass of the inflaton field, up to energy $\Delta^{-1}=\gamma M_\phi$,
with $\gamma$ the relativistic Lorentz factor.

Suppose for illustration that the \WIMPZILLA\ is a fermion coupled to
the inflaton field by a Yukawa coupling $g \phi\overline{X}{X}$. One
can treat $\phi$ (the bubbles or walls) as a classical, external field
and the \WIMPZILLA\ as a quantum field in the presence of this source.
The number of \WIMPZILLAS\ created in the collisions from the wall
potential energy is $N_X\sim f_X M_P/M_X$, where $f_X$ parametrizes
the fraction of the primary decay products in \WIMPZILLAS.  The
fraction $f_X$ will depend in general on the masses and the couplings
of a particular theory in question.  For the Yukawa coupling $g$, it
is $ f_X \simeq g^2 {\rm ln}\left(\gamma M_\phi/2 M_{X}\right)$
\cite{wat,mas}.  \WIMPZILLAS\ may be produced in bubble collisions out
of equilibrium and never attain chemical equilibrium. Even with
$T_{RH}$ as low as 100 GeV, the present \WIMPZILLA\ abundance would be
$\Omega_{X}\sim 1$ if $g\sim 10^{-5}\alpha^{1/2}$.  Here
$\alpha^{-1}\ll 1$ is the fraction of the bubble energy at nucleation
in the form of potential energy at the time of collision.  This simple
analysis indicates that the correct magnitude for the abundance of
\WIMPZILLAS\ may be naturally obtained in the process of reheating in
theories where inflation is terminated by bubble nucleation.

\subsubsection{Wimpzilla Conclusions}

In this talk I have pointed out several ways to generate nonthermal
dark matter.  All of the methods can result in dark matter much more
massive than the feeble little weak-scale mass thermal relics.  The
nonthermal dark matter may be as massive as the GUT scale, truly in
the \WIMPZILLA\ range.  

The mass scale of the \WIMPZILLAS\ is determined by the mass scale of
inflation, more exactly, the expansion rate of the universe at the end
of inflation.  For large-field inflation models, that mass scale is of
order $10^{13}$GeV.  For small-field inflation models, it may be less,
perhaps much less.

The mass scale of inflation may one day be measured!  In addition to
scalar density perturbations, tensor perturbations are produced in
inflation.  The tensor perturbations are directly proportional to the
expansion rate during inflation, so determination of a tensor
contribution to cosmic background radiation temperature fluctuations
would give the value of the expansion rate of the universe during
inflation and set the scale for the mass of the \WIMPZILLA.

Undoubtedly, other methods for \WIMPZILLA\ production will be developed.
But perhaps even with the present scenarios one should start to
investigate methods for \WIMPZILLA\ detection.  While wimpy \WIMPS\ must
be color singlets and electrically neutral, \WIMPZILLAS\ may be endowed
with color and electric charge.  This should open new avenues for
detection and exclusion of \WIMPZILLAS.
The lesson here is that \WIMPZILLAS\ may surprise and be the dark matter, 
and we may learn that size does matter!

\section*{Acknowledgements}

This work was supported by the DOE and NASA under Grant NAG5-7092.
This is an expanded version of my paper ``Early-Universe Issues: Seeds
of Structure and Dark Matter,'' which was part of the proceedings of
the Nobel Symposium {\it Particles and the Universe.}


\newpage

\end{document}